\documentclass[11pt]{article}
\usepackage{amssymb}
\usepackage{amsfonts}
\usepackage{graphics}
\usepackage{amsmath}
\usepackage{color}
\usepackage{graphicx}
\usepackage{caption}
\usepackage{subcaption}
\usepackage{pdflscape}
\usepackage[official]{eurosym}
\usepackage{array}
\usepackage{natbib}
\usepackage[hyphens]{url}
\usepackage[title]{appendix} 
\usepackage[flushleft]{threeparttable} 
\usepackage{hyperref} 
\usepackage[]{float}
\usepackage{comment}

\usepackage{booktabs}
\usepackage{tabularx}

\usepackage[utf8]{inputenc}

\newcounter{coro}
\newcounter{def}
\newcounter{prop}
\newcounter{lem}
\newcounter{probl}
\newcounter{ex}
\newcounter{assum}
\refstepcounter{figure}

\refstepcounter{table}

\newcommand{\sfrac}[2]{\ensuremath\raisebox{1.5pt}{\footnotesize
		$#1$}\kern-1pt/\kern-1pt
	\raisebox{-2pt}{\footnotesize $#2$}}
\def\leaderfill{\leaders\hbox to 1em{\hss.\hss}\hfill}
\def\smallskip{\vskip\smallskipamount}
\def\medskip{\vskip\medskipamount}
\def\bigskip{\vskip\bigskipamount}
\begin{document}
	
\title{The direct and spillover effects of a nationwide socio-emotional learning program for disruptive students.\footnote{We thank the \textit{Junta Nacional de Auxilio Escolar y Becas} (JUNAEB) for allowing us to randomize the timing of the \textit{Habilidades para la vida} program, for providing us with some of the data used in this research, and for answering all our questions about the program. The authors gratefully acknowledge financial support from the Centre for Competitive Advantage in the Global Economy at Warwick University, and from the Economics department at Warwick University. We would also like to thank Juan Pablo Arias, Antonio Figueroa and Gerardo Alvarez for outstanding research assistance. Finally, we also thank Miya Barnett, Myriam George, Peter Kuhn, Loreto Leiva-Bahamondes, Zoe Liberman, Shelly Lundberg, Michael Murphy, Kyle Ratner, Fabian Waldinger, and seminar participants at INSEAD, UC Santa Barbara, University of Bologna, and University of Virginia for their useful comments. This research has been approved by the University of Warwick Research Ethics Committee (approval date: 2014-11-03, approval number: 111/13-14), and has been registered on the social science registry website (RCT ID AEARCTR-0001080). $\dagger$ Cl\'ement de Chaisemartin: UC Santa Barbara; J-Pal; NBER. $\ddagger$Nicol\'as Navarrete H: Paris School of Economics. For questions please email: \texttt{clementdechaisemartin@ucsb.edu} or \texttt{nicolas.navarrete@psemail.eu}.}}
\author{Cl\'ement de Chaisemartin$\dagger$ \and Nicol\'as Navarrete H.$\ddagger$}
\maketitle

\begin{abstract}
Social and emotional learning (SEL) programs teach disruptive students to improve their classroom behavior. Small-scale programs in high-income countries have been shown to improve treated students’ behavior and academic outcomes. Using a randomized experiment, we show that a nationwide SEL program in Chile has no effect on eligible students. We find evidence that very disruptive students may hamper the program's effectiveness. ADHD, a disorder correlated with disruptiveness, is much more prevalent in Chile than in high-income countries, so very disruptive students may be more present in Chile than in the contexts where SEL programs have been shown to work.
\end{abstract}

\textnormal{Keywords:} disruptive students, spillover effects, peer effects, social and emotional learning.

\medskip
\textbf{JEL Codes:} I21, I24, I28, D62.

\let\markeverypar\everypar
\newtoks\everypar
\everypar\markeverypar
\markeverypar{\the\everypar\looseness=-2\relax}
\thispagestyle{empty}
\parskip=3.pt
\newpage
\baselineskip=20pt
\linespread{1}


\section{Introduction}
\label{sec:Introduction}
\setcounter{page}{1}

\cite{lazear2001} has proposed that classroom learning is a public good suffering from congestion effects, which are negative externalities created
when one student is disruptive and impedes the learning of her classmates. In the US, those externalities are important: \cite{carrell2010externalities} and \cite{carrell2018long} find that being exposed to one peer experiencing domestic violence at home, a good proxy for a disruptive peer, reduces classmates' test scores by 0.07 standard deviation ($\sigma$), and reduces their earnings at age 26 by 3 to 4 percent. \cite{figlio2007} also finds that being exposed to disruptive peers reduces classmates test scores. \cite{betts1999} find that US middle and high schools teachers devote 6.1\% of instruction time to discipline, and that this fraction is higher in disadvantaged schools. Therefore, programs effective at reducing troubled students' disruptiveness may generate large positive spillover on their classmates, on top of their direct effects. 

\medskip
Epidemiological studies show that the prevalence of ADHD, a disorder correlated with conduct problems, is higher in some low- and middle-income countries than in high-income countries. Then, addressing students' conduct problems may be an even more pressing issue in those countries. In Chile, the country where the intervention we study takes place, 15.5\% of primary school children have ADHD (see \citep{de2013epidemiology}). Primary school children also have be found to have high ADHD rates in Colombia (16.9\%, see \citep{cornejo2005prevalencia}), or in Iran (17.3\%, see \citep{safavi2016prevalence}).  On the other hand, the ADHD prevalence rate among primary school children is estimated at 6.8\% in the US (see \citep{visser2014trends}), between 3.5 and 5.6\% in France (see \citep{lecendreux2011prevalence}), and 3\% in Italy (see \citep{bianchini2013prevalence}).

\medskip
School-based mental health programs are often used to reduce students' disruptiveness.
Some programs are universal, meaning that they are delivered
in classroom settings to all the students in the
class. Other programs are selected, meaning that they are provided to students identified by teachers as having conduct problems, during the school day and outside the classroom.
Many school-based mental health programs are social and emotional learning (SEL) programs (see \citep{wilson2007school}), that teach children to recognize and manage their emotions, and to handle interpersonal situations effectively, using cognitive and behavioral therapy (CBT). A vast literature has found SEL programs to be successful. In a meta-analysis of 80 selected interventions, \cite{payton2008} find that they reduce conduct problems by $0.47\sigma$, and respectively improve mental health and academic performance by $0.50$ and $0.43\sigma$. Meta-analyses of universal SEL interventions find smaller but still large effects on those dimensions, around $-0.25\sigma$ for conduct problems, and $+0.55\sigma$ and $+0.30\sigma$ for mental health and test scores (\citep{durlak2011impact}, \citep{sklad2012effectiveness}, \citep{wigelsworth2016impact}, \citep{taylor2017promoting}, and \citep{corcoran2018effective}).

\medskip
However, there are at least two gaps in the literature. First, it has mostly considered small-scale demonstration programs mounted by researchers in a handful of schools. The effect of SEL interventions may differ when implemented at scale (see \citep{davis2017} or \citep{weisz2014}). Second, it has mostly focused on interventions conducted in high-income countries, while epidemiological studies suggest that addressing students' conduct problems may be more pressing in some middle- and low-income countries.
A recent meta-analysis of psycho-social interventions for disruptive students in low- and middle-income countries (see \citep{burkey2018psychosocial}) includes only two SEL interventions, one in Jamaica and the other in Romania. Both are universal interventions, implemented at a very small scale. Both find positive effects on students' conduct problems (see \citep{baker2009pilot} and \citep{cstefan2013}).

\medskip
This paper contributes to addressing those two gaps: it is the first to measure the effects of a nationwide SEL program, and the program we study takes place in a middle-income country with a high ADHD prevalence rate. Specifically, we study the effects of ``Skills for Life'' (SFL), a selected SEL program for disruptive second graders in Chile. Since its creation in 1998, SFL has screened and treated around 1,000,000 children, making it the fifth largest school-based mental health program in the world (see \citep{murphy2017}). To identify eligible students, SFL teams use a psychometric scale measuring students' disruptiveness, and students above some cut-off are eligible. Eligible students then follow 10 two-hours SEL sessions with a psychologist and a social worker. SFL is a costly program: we estimate that its cost per student is equivalent to 15\% of the expenditure per primary school student in Chile.

\medskip
We randomly assigned 172 classes to either receive SFL in the first or second semester of the 2015 school year, and we measured outcomes at the start of the second semester, after the treatment group had received the treatment but before the control group received it. By comparing eligible students in the treatment and control groups, we can estimate the direct effects of the program, and by comparing ineligible students in the two groups we can estimate its spillover effects.

\medskip
We find that SFL does not have effects on eligible students' disruptiveness, mental health, and academic achievement. The effects we can rule out are fairly small, and much smaller than those found in \cite{payton2008} and in all the other SEL meta-analyses we are aware of. For instance, we can rule out at the 5\% level that the program increases students’ Spanish scores by more than 0.09$\sigma$, or that it reduces teachers’ assessment of students’ disruptiveness by more than 0.10$\sigma$. Not surprisingly, as we do not find that SFL impacts eligible students, we also do not find spillover effects on ineligible ones. Finally, we even find that the program has a strong negative effect on teachers' and enumerators' ratings of the overall disruptiveness of treated classes.

\medskip
To account for the discrepancy between our results and the literature, we compared SFL to the selected SEL interventions reviewed in \cite{payton2008}. Three conclusions emerge.
First, SFL's intensity (number of sessions, duration...) is comparable to that of the meta-analysis's interventions, so it is not the case that SFL is not intensive enough to produce an effect. Second, as ADHD is much more prevalent in Chile than in high-income countries, SFL may be faced with a harder-to-treat population than the interventions reviewed in \cite{payton2008}.
We indeed find evidence that SFL's effectiveness is hampered by the presence of very disruptive students. In classes with at least one very disruptive eligible student, defined as students above eligible students' 90th percentile of baseline disruptiveness,\footnote{This definition ensures that about 50\% of classes have at least one very disruptive student.} the program increases the disruptiveness of other eligible students, of ineligible students, and it worsens teachers' and enumerators' ratings of the overall class disruptiveness. The program also strongly increases the friendship ties between very disruptive and other eligible students. The former may then have a negative influence on the latter, which would explain the negative effects we observe. Third, SFL and the meta-analysis's interventions strikingly differ in terms of scale and delivery.
The interventions in the meta-analysis are demonstration programs mounted by researchers, that typically treat a few dozens children in a handful of schools.
Half are delivered by the researchers, while the other half are delivered by psychologists or teachers under researchers' close supervision: typically, researchers review their delivery of the intervention every week. On the other hand, SFL is a large-scale governmental program, delivered by psychologists without any researcher involvement. The governmental agency in charge of the program loosely monitors the program implementers, and very rarely audits their workshops. Without sufficient monitoring, teams may not implement the program with high-enough fidelity, which could also explain why SFL does not produce an effect.

\medskip
The remainder of the paper is organized as follows. In Section \ref{sec:treatment}, we present the SFL program. In Section  \ref{sec:studydesign}, we present the  randomization, the data we use, and the population under study. In Section \ref{sec:Compliance_internalvalidity_estimation}, we present compliance with randomization, the balancing checks, and attrition. In Section \ref{sec:Treatment Effects}, we present the main results. In Section  \ref{sec:mech}, we interpret the results and present some exploratory analysis.

\section{The SFL program}
\label{sec:treatment}

SEL is the process through which children acquire the skills to recognize and manage their emotions, set and achieve positive goals, and handle interpersonal situations effectively. SEL programs try to enhance children's self-awareness (accurately assessing one's feelings and maintaining a sense of self-confidence), self-management (regulating one's emotions and controlling impulses), and social awareness (being able to take the perspective of others, preventing, managing, and resolving interpersonal conflict). 
Selected programs are provided to specific students identified as having conduct problems, during the school day and outside of their classroom. Meta-analysis have shown that selected SEL programs improve SEL skills, reduce conduct problems, and can improve academic achievement (see \citep{payton2008}).

\medskip
SFL is a Chilean school-based selected SEL program for second graders suffering from conduct disorders. It is managed by JUNAEB (\textit{Junta Nacional de Auxilio Escolar y Becas}), the division of the Chilean Department of Education in charge of most of the non-teaching programs implemented in Chilean schools. The program started as a pilot in 1998. Over the next 3 years, JUNAEB collaborated with psychologists from the University of Chile 
to review the screening measures and programs available at that time, and design an SEL program adapted to Chile, where the prevalence of ADHD is particularly high among children (see \citep{de2013epidemiology}). The program became a nationwide policy in 2001, and it is currently implemented in 1,637 publicly-funded elementary schools in Chile (see \citep{guzman2015}). These schools account for 20\% of all elementary schools in Chile, and they are the most disadvantaged.
Since 1998, the SFL program has screened and treated around 1,000,000 children, making it the fifth largest school-based mental health program in the world (see \citep{murphy2017}).

\medskip
To identify eligible students, SFL uses a psychometric scale, the Teacher Observation of Classroom Adaptation (TOCA, see \citep{kellam1977}, and \citep{werthamer1990}), adapted to the Chilean context by \cite{george1994adaptacion}. In the end of each academic year, first-grade teachers answer the TOCA questionnaire for each of their student. Based on this
questionnaire, students receive scores on the following six scales: authority acceptance (AA), attention and focus (AF), activity levels (AL), social
contact (SC), motivation for schooling (MS), and emotional maturity (EM). The TOCA questionnaire concludes with two summary questions, where teachers have to give ratings of the overall disruptiveness and academic ability of each of their student.

\medskip
Then, the three following groups of students are eligible for the program:
\begin{itemize}
\item Students above the 75th percentile of the AA scale, above the 85th percentile of the AF and AL scales, and below the 25th percentile of the MS scale;
\item Students below the 25th percentile of the SC scale, and either above  the 75th percentile of the AA scale or above the 85th percentile of the AL scale;
\item Students below the 25th percentile of the SC, MS, and EM scales, and below the 50th percentile of either the AA or AL scale.
\end{itemize}
The percentiles are gender specific, to ensure that not only males are eligible, and were computed using a representative sample of the 2nd grade population in Chile (see \citep{george1994adaptacion,de2005prediction}).
Students in the third eligibility group are not disruptive, but they only account for 7\% of eligible students, while the first two groups respectively account for 40\% and 53\% of eligible students. Depending on the year, eligible students account for 15 to 20\% of first-grade students whose teachers answer the TOCA questionnaire.

\medskip
In second grade, SFL asks eligible students' parents the authorization to enroll their child in the program. If their parents accept, eligible students are enrolled in a workshop implemented by a team of two SFL employees. A survey conducted in 2015 (see \citep{rojas2018}) shows that half of SFL employees are psychologists. In Chile, this title can be obtained after a college degree with a psychology major (see \citep{guzman2015}). The other half of employees are social workers and former teachers, titles that can also be obtained after a college degree. Usually, an SFL team consists of a psychologist and a social worker or teacher. 77\% of SFL employees are women, their average age is 31 years old. They have on average 2.6 years of experience into the program, and 36\% have less than one year of experience, indicating a high rate of turnover. During their first year, SFL employees receive three eight-hours-long days of training (see \citep{rojas2017efectos}). They also attend ``good practices'' meetings every six months, in which they share with other teams what works in their workshops. As the Chilean public school system is administrated at the municipal level, SFL teams are also organized at this administrative level. 

\medskip
SFL workshops consist in 10 two-hours group sessions,
taking place weekly, during the class day, over the course of one semester. During sessions, enrolled students leave the classroom, while their classmates stay there and continue with their normal schedule. The time of the group sessions is set in coordination with teachers, to avoid that enrolled students lose key instruction time. During the workshops, teachers teach subjects deemed less crucial than Spanish or mathematics, like religion (a mandatory subject in Chile) or music, to the ineligible students that stay with them in the classroom (see \citep{rojas2018}). The workshop takes place over two school periods, and eligible students come back to their classroom before the break.

\medskip
Sessions are divided into five parts. The goals of the first part are to welcome children and build a group identity, for instance by having children choose a group name. The goals of the second part are to improve children's self-esteem, and their respect of others. Then, during the third part, the psychologists help students put words on their and others' emotions, and help them share their emotions with others. Then, the fourth part is dedicated to self-control techniques, and to strategies to find non-violent solutions to conflicts. Finally, the last part is dedicated to a review of what has been learnt during the workshop. Sessions are activity based, involve games and role play, and make use of CBT techniques. If they behave well during a session, students sometimes receive rewards like cakes or candies. SFL employees are provided with a 114-pages-long manual describing the goal and the content of each session, and suggesting games and activities. But they are also encouraged to tailor the content of their sessions to the specific needs of the students enrolled.

\medskip
As per the SFL guidelines, six to 12 students should participate in a workshop. If there are less than six eligible students in a school, no workshop takes place, and if a school has more than 12 eligible students, two workshops take place in that school. In the next section, we explain how we exploit these features in our randomization. Finally, the parents of enrolled children are invited to three training sessions, whose goal is to  encourage them to reproduce the workshop's activities at home.

\medskip
We estimate that SFL costs 200 USD per treated student. We also estimate that the government spends 1,316 USD on instruction per student and per year in the schools in our sample.\footnote{The government funds public schools by giving them a voucher per student, whose amount depends on the student's attendance (\url{https://www.oecd-ilibrary.org/docserver/9789264287112-6-es.pdf?expires=1586606397\&id=id\&accname=guest\&checksum=17AF0B3C9CF0863F8300FAA082FE969D}). For public primary schools, the school voucher is worth 754 USD for an attendance of 84\%, the average attendance observed  in our sample. Then, the government gives schools an extra voucher worth 721 USD for every very disadvantaged student (\url{https://ate.mineduc.cl/usuarios/admin3/doc/2015020312570909985.Manual_Apoyo_a_la_Gestion.pdf}), and 78\% of students are very disadvantaged in the schools we study, thus leading to our 754+0.78$\times$721=1,316 USD estimate.} Therefore, the program's cost represents a sizeable 15\% increase of the expenditure per student. JUNAEB does not have an estimate of the total cost of the program, here is how we estimated it. The 2014 budget of one of the municipal teams in our sample shows that its program implementers earned on average 7.42 USD per hour in 2014. Then, based on interviews with two implementers, we estimated that it takes 149 hours of work to implement an SFL workshop. This includes the 52 hours that the two workshop implementers spend delivering 13 two-hours sessions to students and their parents, but also the time that they spend: preparing the sessions and buying the material they need; going to and returning from the school for each session; preparing the reporting documents JUNAEB asks them to send for each workshop; meeting with the school principal and 2nd grade teachers prior to the start of the workshop, to agree on the schedule and location of the workshop; and interview 1st grade teachers to fill the TOCA questionnaire for each of their students the year before the workshop. Then, the team's budget shows that digitizing the 2014 TOCAs of all the first grade students in the town costed 860 USD. Divided by the 10 workshops conducted that year, that leads to a cost of 86 USD per workshop. Implementers also received transportation vouchers worth 63 USD per workshop. Finally, the cost of the material needed for the workshop activities is estimated at 188 USD per workshop, based on a detailed list of all the items bought for a workshop provided by the implementers we interviewed. Overall, we estimate the total cost of a workshop at 7.42$\times$149+86+63+188=1,443 USD. The team whose budget we used had 7.2 students per workshop in 2014, which finally yields our estimated cost of 200 USD per treated student. This estimate relies on one team's budget. Costs may vary between teams, but we do not have reasons to suspect that the program's average cost is orders of magnitude away from our estimate. 

\medskip
Previous research has found that from first to third grade, the disruptiveness of students that attend seven to 10 SFL sessions in second grade decreases more than that of students attending six sessions or less (see e.g. \citep{guzman2015}). However, SFL attendance is driven by students' school attendance, and students who attend school less may do so because they experience negative shocks, which could explain why their disruptiveness decreases less. To avoid that type of endogeneity bias, our paper relies on an experimental control group to measure the effect of SFL. 

\section{Randomization, data, and study population}
\label{sec:studydesign}

\subsection{Sample selection and randomization}\label{sec:design}

Our sample consists of 172 classes. All municipal teams conducting the SFL program in the Santiago and Valparaiso regions, the two most populated regions in Chile, were invited to join the study. 32 out of 39 accepted our invitation. In March 2015, these teams visited the schools covered by the program in their municipalities, and collected data on the number of students eligible for the program enrolled in each second grade class. 172 classes with four or more eligible students and in schools with six or more eligible students were included in the study. The second criterion ensured that group sessions would indeed take place in the school, while the first criterion ensured that there were enough treated students per class to potentially generate spillover effects. About 450 classes participate in a SFL workshop each year in the Santiago and Valparaiso regions, so our sample covers about 40\% of the classes covered by the program in those regions.

\medskip
Randomization took place both within schools and within municipalities. There were 29 schools with two classes included in our sample and where it was possible to form two groups of six students or more without grouping students of the two classes together. In such instances, we conducted a lottery within the school, to assign one of the two classes to receive the treatment in the first semester of 2015, and the second class to receive it in the second semester. The remaining 114 schools each only had one class included in our sample, so randomization took place within municipalities. In this latter group of schools, there is no risk that the control group students may have been contaminated by the treatment, while this may have happened in the former group of schools. Later in the paper, we reestimate the treatment effect in the second group of schools and find very similar effects to those we find in the full sample, so control-group contamination does not seem to drive our results.

\medskip
Overall, we conducted 56 lotteries (29 within schools, and 27 within municipalities) and we assigned 89 classes to receive the treatment in the first semester, from April to June 2015, and 83 to receive it in the second semester, from September to December 2015.

\subsection{Data}\label{sec:data}

In our analysis, we use data produced by JUNAEB. First, we use the six first-grade TOCA scores that determine students' eligibility to SFL, as well as the teachers' ratings of students' disruptiveness and academic ability in the TOCA questionnaire. Then, we also use another psychometric scale collected by JUNAEB and measuring students' disruptiveness, the pediatric symptom checklist (PSC, see \citep{jellinek1988}), which is filled by students' parents. We also use JUNAEB's data on treatment implementation. Specifically, for each class in our sample we know how many SFL group sessions were conducted in the first semester of 2015. For each student, we know how many sessions she attended, and how many sessions her parents attended. Finally, JUNAEB also provided us data on students' socio-economic background, as well as their monthly school attendance from March 2015 to June 2015.

We also use baseline data collected in March 2015, before the treatment started in the treatment group classes, and endline data collected in August 2015, after the treatment ended in the treatment group classes and before it started in the control group classes. Both at baseline and endline, two enumerators visited each of the 172 classes included in the experiment during a half day. Enumerators were undergraduate students, mostly psychology and education majors. Every person who applied to become an enumerator first had to attend a half-day training, during which he/she was taught how to administer our questionnaires. Candidates also had to take a test at the end of the training, and only those who scored above some threshold became enumerators.

\medskip
Our questionnaires slightly changed from baseline to endline. Below, we describe our endline questionnaires, and we explain the difference between our baseline and endline questionnaires when needed later in the paper.

\medskip
The enumerators first administered a non-cognitive questionnaire to the students. That questionnaire aimed at measuring:
\begin{itemize}
\item Students' happiness in school, using a question from the student SIMCE questionnaire.\footnote{The SIMCE (\textit{Sistema de Medici{\'o}n de la Calidad de la Educaci{\'o}n}) questionnaires are the nationwide standardized cognitive and non cognitive questionnaires administered to students and teachers in Chile.}
\item Students' self-control, using items of the child self-control psychometric scale (see \citep{rorhbeck1991child}) that we translated into Spanish.
\item Students' self-esteem, using items of the self-perception for children psychometric scale (see \citep{harter1985manual}) translated and validated into Spanish (see \citep{molina2011adaptacion}).
\end{itemize}

\medskip
Second, the enumerators administered a Spanish and mathematics test to the students. Third, the enumerators interviewed individually each student and asked her to name up to three students that she likes to play with during breaks, hereafter referred to as the student's friends. Fourth, the enumerators observed a one-hour lecture. During that observation, they observed the behaviour of each student during five seconds, and assessed whether the student was studying, not studying, or being disruptive. They repeated that process five times, and then rated the overall disruptiveness of each student by answering the summary question from the TOCA questionnaire. During that one-hour lecture, the enumerators also recorded the decibel levels in the class using a smartphone app, and wrote down the time at which the lecture was supposed to start and the time when it effectively started. Fifth, the enumerators filled a short questionnaire aimed at assessing the overall disruptiveness in the class, using questions taken from the PISA (Program for International Student Assessment) questionnaire, asking them their agreement with statements such as: ``There is noise and disorder in this class,'' or ``The teacher has to wait for a long time before students calm down and he/she can start teaching''.

\medskip
The enumerators also administered a questionnaire to the teachers. That questionnaire aimed at collecting: teachers' socio-demographic characteristics; teachers' ratings of the overall disruptiveness of the class, using similar questions as those asked to enumerators; teachers' rating of the prevalence of bullying in the class;
teachers' motivation, taste for their job, and mental health levels. The questionnaire was for the most part composed of questions from the SIMCE teacher questionnaire. Teachers also rated the overall disruptiveness of each of their student by answering the summary question from the TOCA questionnaire.

\medskip
Finally, in July 2019 we also conducted qualitative interviews to shed light on the mechanisms underlying our results. We interviewed three of the SFL municipal teams that had participated in our experiment, and that account for 12\% of our sample.

\medskip
The list of the outcome variables we consider in the paper was pre-specified in a pre-analysis plan (PAP) available at \url{https://www.socialscienceregistry.org/trials/1080}. That plan was time-stamped on 04/28/2017, before JUNAEB sent us students' first grade TOCA scores, as a letter from JUNAEB officials also available on the social science registry website testifies. Students' first-grade TOCA scores are necessary to distinguish eligible and ineligible students in our data, a distinction that underlies most of our analysis. Even though endline took place almost two years before we submitted our PAP, we had not started to analyze our data before. 
Indeed, we had not finished cleaning the data before submitting our PAP. This research was funded through four small grants, totalling 37,000 GBP. Therefore, we could not afford to buy tablets to collect our data, and instead used paper questionnaires. We could also not afford to hire a RA in charge of supervising data entry and data cleaning. Instead, we supervised the RAs in charge of data entry ourselves, and we also took care of the data cleaning ourselves. This process ended after 04/28/2017.

\medskip
The analysis presented in Sections \ref{sec:Compliance_internalvalidity_estimation} and \ref{sec:Treatment Effects} follows our pre-analysis plan, except for a few exceptions described below. On the other hand, the analysis presented in Section \ref{sec:mech} was not pre-specified in our PAP. The student-level outcome measures listed in our PAP are:
\begin{itemize}
\item the student's happiness in school, self-control, self-esteem, Spanish, and mathematics scores,
\item the percentage of school days missed by the student from April to June 2015,
\item the rating of the student's disruptiveness by her teacher,
\item the average rating of the student's disruptiveness across the two enumerators,
\item the percentage of the student's classmates that nominate her as one of their friends,
\item an indicator for whether the student is not nominated as a friend by any other student,
\item the average disruptiveness at baseline of the student's endline friends,
\item the average baseline Spanish and mathematics scores of the student's endline friends.
\end{itemize}
The class-level outcome measures listed in our PAP are:
\begin{itemize}
\item the teacher's rating of the class's disruptiveness, constructed using teachers' answers to the PISA questions measuring the disruptiveness in the class,
\item the teacher's rating of the prevalence of bullying in the class,
\item the average rating of the class's disruptiveness across the two enumerators, constructed using enumerators' answers to the PISA questions measuring the disruptiveness in the class,
\item the number of minutes between the moment the class was supposed to start and the moment it effectively started according to the enumerators,
\item the average decibel levels during the class across the two enumerators' recordings.
\end{itemize}
We standardize the school happiness, self-control, self-esteem, disruptiveness and test score measures to have a mean of 0 and a $\sigma$ of 1 in the sample.

\subsection{Assessing data quality}\label{sec:data_quality}

Some of the dimensions we are trying to measure are hard to observe. To get a sense of the reliability of our measures, Table \ref{tab:corr_1} shows their baseline-endline correlation in the control group. Students' Spanish and mathematics test scores have high positive baseline-endline correlations, above 0.5. Those correlations are still far from one, probably because students in our study are young and their cognitive ability is not fixed yet. Our measure of students' popularity has a baseline-endline correlation of 0.32. Our school happiness, self-esteem, and self-control measures respectively have baseline-endline correlations of 0.22, 0.13, and 0.14.

\medskip
Turning to disruptiveness measures, the rating of students' disruptiveness by teachers has a baseline-endline correlation of 0.42, which is almost as high as the baseline-endline correlation of test scores. This is all the more remarkable as we use first grade teachers' answer to the TOCA summary question as our baseline measure,\footnote{We decided to include the summary TOCA question in our baseline teacher questionnaire after having collected more than half of the baseline data, so that variable is missing for many classes at baseline.} so our baseline and endline measures were not made by the same teacher. This suggests that students' disruptiveness is relatively stable, and that different teachers tend to agree in their ratings.
Then, Table \ref{tab:corr_2} shows that this measure is negatively correlated with students' academic ability: at baseline, its correlation with students' average test score in Spanish and mathematics is equal to -0.28. Finally, the bottom panel of Table \ref{tab:corr_1} shows that teachers' rating of the disruptiveness of the class also has a high baseline-endline correlation, equal to 0.50.

\medskip
In our PAP, we had planned to use the average of the two enumerators' ratings of a student's disruptiveness as our enumerator disruptiveness rating. However, this measure has a baseline-endline correlation close to, and insignificantly different from, zero. This could be due to the fact that endline and baseline observations are made by different enumerators, who may have different standards to assign a given grade on the disruptiveness scale. Therefore, we depart from our PAP, and slightly modify our measure. We start by regressing enumerators' ratings on enumerator fixed effects, in the sample of control group classes. Then, we compute the residuals from that regression both for treatment and control group classes, and we use the average of those residuals, across the two enumerators that have rated a student, as our enumerators' rating. This modified measure is the difference between a student's average rating by the two enumerators and the average of the ratings made by the same enumerators in the control group. Panel A of Table \ref{tab:corr_1} shows that it has a positive and significant baseline-endline correlation equal to 0.13, and Panel A of Table \ref{tab:corr_2} shows that it correlates well with teachers' ratings, and reasonably well with students' academic ability. Overall, enumerators' ratings of students' disruptiveness seem noisier than teachers', but they are still meaningful.
Then, Panel B of Table \ref{tab:corr_1} shows that enumerators' ratings of classes' disruptiveness have a relatively high baseline-endline correlation, around 0.25, and Panel B of Table \ref{tab:corr_2} shows that this measure correlates well with teachers' ratings.
Contrary to teachers' ratings, enumerators' ratings are blinded: enumerators do not know if the class they observe has been treated or not.\footnote{Previous literature on SEL interventions has also relied on non-blinded teacher ratings (see \citep{payton2008}).}

\medskip
The decibel measure constructed following our PAP also has a very low baseline-endline correlation, and it does not correlate at all with teachers' and enumerators' ratings of classes' disruptiveness. The app's measurement does not seem very precise: enumerators recording the same lecture sometimes end up with average noise levels differing by more than 10 decibels. This measurement also seems to depend on the make of the phone and on idiosyncratic factors specific to the enumerator's phone. Therefore, we depart again from our PAP, and net out enumerators' fixed effects from decibel measures, exactly as we did for enumerators' disruptiveness ratings. This new measure has a higher baseline-endline correlation than the measure described in our PAP, though Table \ref{tab:corr_1} shows that this correlation is still not significant. But it also has a much larger correlation with enumerators' ratings of the class disruptiveness, and that correlation is significant as shown in Table \ref{tab:corr_2}.


\subsection{Study population}\label{sec:applicants}

The 172 classes included in our sample bear 5,704 students, meaning that classes have an average of 33.2 students. 4,466 students are ineligible to the program (26.0 per class), while 1,238 students are eligible (7.2 per class). Column (1) in Table \ref{tab:statdes1} below presents the baseline characteristics of ineligible students. 33.8\% of them are born to teenage mothers, which is more than twice the corresponding proportion in Chile.\footnote{See \url{http://web.minsal.cl/portal/url/item/c908a2010f2e7dafe040010164010db3.pdf}.} 75.2\% of them live in households below the 20th percentile of the social security score. Being below this threshold opens eligibility for 22 social programs and is usually considered as a proxy for poverty. 44.4\% of them live in households below the 5th percentile of the social security score. Being below this threshold opens eligibility for 3 more social programs and is usually considered as a proxy for extreme poverty. Overall, the students included in our study live in households disproportionately coming from the bottom of the Chilean income distribution.

\medskip
Column (2) in Table \ref{tab:statdes1} presents the baseline characteristics of eligible students, and Column (3) reports the p-value of tests that the baseline characteristics of eligible and ineligible students are equal. Panel A shows that eligible students are more likely to be males and less likely to live with their father. Their parents are also less educated than that of ineligible students. Panel B shows that eligible students's self-control and self-esteem scores are about 0.2$\sigma$ lower than that of ineligible students. Differences are even more pronounced when one considers students' disruptiveness and academic ability. Eligible students score 1.2$\sigma$ higher than ineligible students on first-grade teachers' disruptiveness ratings, and 0.4$\sigma$ higher on enumerators' baseline ratings. They also score 0.4$\sigma$ lower on the Spanish and mathematics tests. Eligible students are also less popular than ineligible ones: 7.6\% of the students in the class nominate them as friends, against 8.8\% for ineligible students. The average disruptiveness of their friends is also about 0.2$\sigma$ higher than that of ineligible's friends, thus suggesting some assortative matching along the disruptiveness dimension.

\medskip
Finally, Table \ref{tab:teachstatdes} shows some characteristics of the teachers in our sample. 96.3\% of teachers are females. Their average age is 42.8 years old, they have an average of 16.5 years of experience as a teacher, and 8.6 years of experience in the school where they currently teach.

\begin{table}[H] \centering
  \begin{threeparttable}
\caption{Characteristics of eligible and ineligible students}
\label{tab:statdes1}
\normalsize

\begin{tabular}{lccccccc}
\toprule\toprule
{}&{Ineligible}&{Eligible}&{P-value}&{N} \tabularnewline
&{(1)}&{(2)}&{(3)}&{(4)} \tabularnewline
\midrule\multicolumn{5}{c}{Panel A: demographic characteristics}\tabularnewline\midrule
Male&0.498&0.582&0.000&5704 \tabularnewline
Teen mother&0.338&0.36&0.199&4440 \tabularnewline
Student lives with father&0.635&0.554&0.000&3765 \tabularnewline
$\leq$ p20 social security score&0.752&0.77&0.198&5068 \tabularnewline
$\leq$ p5 social security score&0.444&0.456&0.469&5068 \tabularnewline
Mother's education&9.131&8.564&0.000&4727 \tabularnewline
Father's education&9.163&8.439&0.000&4117 \tabularnewline
\midrule\multicolumn{5}{c}{Panel B: baseline measures}\tabularnewline\midrule
School happiness score&0.023&-0.063&0.022&4431 \tabularnewline
Self-control score&0.048&-0.166&0.000&4594 \tabularnewline
Self-esteem score&0.041&-0.146&0.000&4610 \tabularnewline
Overall disruptiveness TOCA&-0.293&0.873&0.000&4850 \tabularnewline
Disruptiveness, enumerator&-0.03&0.341&0.000&4645 \tabularnewline
Spanish test score&0.095&-0.335&0.000&4758 \tabularnewline
Math test score&0.082&-0.289&0.000&4758 \tabularnewline
\% class friends with student&0.088&0.076&0.000&4721 \tabularnewline
Friends' average disruptiveness&-0.051&0.188&0.000&3931 \tabularnewline
\bottomrule
\end{tabular}

  \begin{tablenotes}[para,flushleft]
\footnotesize {\it Notes:} This table reports descriptive statistics for students in the sample. Column (1) reports the mean of the outcome variable for ineligible students and Column (2) reports the mean of the outcome variable for eligible students. Column (3) reports the p-value of a test that the two means are equal. Column (4) reports the number of observations used in the comparison.

\end{tablenotes}
  \end{threeparttable}
\end{table}

\section{Compliance, internal validity, and estimation methods}
\label{sec:Compliance_internalvalidity_estimation}

\subsection{Compliance with randomization and fidelity of treatment assignment}
\label{subsec:Compliance}

In this section, we show that the SFL teams followed the randomization, and implemented the treatment as per the program's rules: in the treatment group classes, very few ineligible students received the program. To do so, we estimate the effect of being assigned to treatment on actual exposure to treatment during the first semester of 2015.
Let $Y_{ijk}$ be a measure of exposure to treatment for student $i$ in class $j$ and lottery $k$. We estimate the following regression:
\begin{equation}\label{eq:regression1}
Y_{ijk}=\gamma_k+\beta D_{jk}+u_{ijk},
\end{equation}
where the $\gamma_k$s are fixed effects for the 56 lotteries we conducted to assign the treatment, and where $D_{jk}$ is equal to 1 if lottery $k$ assigned class $j$ to the treatment group and to 0 otherwise. $\widehat{\beta}$ estimates a weighted average across lotteries of the within-lottery difference between the average of $Y_{ijk}$ in treatment and control group classes. As our lotteries have few classes, the treatments of classes in the same lottery are strongly negatively correlated. Therefore, we cluster standard errors at the lottery level, following the recommendation of \citep{Chaisemartin2019clusterpairRCTs}), who show that clustering at the class level could lead to substantial over-rejection of the null hypothesis.

\medskip
To estimate the effect of assignment to treatment on class-level measures of exposure, we estimate Regression \eqref{eq:regression1}, except that we use propensity score reweighting instead of lottery fixed effects. With propensity score reweighting, $\beta$ is also identified out of comparisons of treatment and control group classes in the same lottery (see \citep{hirano2003}). Using propensity score reweighting ensures that the regression does not have too many independent variables with respect to its number of observations (with lottery fixed effects, Regression \eqref{eq:regression1} would have 57 independent variables and at most 172 observations). In any case, as the share of treated classes is equal to 0.5 in 46 of the 56 lotteries, using lottery fixed effects or propensity score reweighting does not make a large difference.

\medskip
Column (1) of Table \ref{tab:H13} below shows the mean value of eight measures of exposure to the treatment in the control group. Column (2) shows estimates of $\beta$ for these eight measures. Column (3) shows estimates of the standard error of $\widehat{\beta}$. Column (4) shows the p-value of a t-test of $\beta=0$. To account for the fact that we consider several measures of exposure to the treatment, Column (5) shows the p-value controlling the False Discovery Rate (FDR) across the eight tests (see \citep{benjamini1995controlling}). Finally, Column (6) shows the number of observations used in the estimation.

\medskip
Panel A of the table shows that SFL sessions were conducted in 8.4\% of the control group classes and in 98.1\% of the treatment group classes. On average, 0.6 sessions were conducted in the control group classes against 9.5 in the treatment group classes. Throughout the paper, we estimate intention to treat (ITT) effects of assigning a class to the treatment. Given that less than 10\% of the control group classes received the treatment, while almost 100\% of treatment group classes received it, this ITT effect ``almost'' estimates the effect of delivering the treatment in a class.

\medskip
Panel A also shows that 4.8\% of eligible students in the control group attended at least one session, against 84.9\% in the treatment group. Some eligible students did not attend any group session, either because their parents refused that they participate, or forgot to send back the document they had to sign to authorize their child's participation. Table \ref{tab:statdes_takersnontakers1} compares the characteristics of the ``takers'', eligible students in the treatment group that attended at least one session, to those of the ``non takers'' that did not attend any session. The main difference between the two groups is that the takers are less disruptive at baseline.
On average, eligible students attended 0.4 sessions in the control group, against 7.4 in the treatment group. This number is 8\% lower than $9.5\times 0.849=8.1$, the number we would have observed if students attending at least one session had attended all the sessions conducted in their class. This small difference is due to the fact that those students sometimes miss school on a workshop day, but school absenteeism does not seem to reduce students' exposure to the program very much.
Finally, Panel A shows that the fidelity with the program's assignment rules was very high: in the treatment group, only 1\% of ineligible students attended at least one session.

\medskip
Panel B of the table shows that compliance with randomization was lower for the parents' than for the students' workshops: 53.5\% of eligible parents in the treatment group attended at least one session, and eligible parents attend on average 1.0 sessions out of 3.

\begin{table}[H] \centering
  \begin{threeparttable}
\caption{Compliance with randomization}
\label{tab:H13}
{\normalsize
\begin{tabular}{lccccccc}
\toprule\toprule
{Variables}&{Control}&{T-C}&{S.E.}&{Unadj. P}&{Adj. P}&{N} \tabularnewline
&{(1)}&{(2)}&{(3)}&{(4)}&{(5)}&{(6)} \tabularnewline
\midrule\multicolumn{7}{c}{Panel A: students' workshops}\tabularnewline\midrule
$\geq$1 session conducted in class&0.084&0.897&0.035&0.000&0.000&172 \tabularnewline
Sessions conducted in class&0.602&8.942&0.337&0.000&0.000&172 \tabularnewline
Eligible students attended $\geq$1 session&0.048&0.801&0.029&0.000&0.000&1238 \tabularnewline
Sessions attended by eligible students&0.37&6.992&0.304&0.000&0.000&1238 \tabularnewline
Ineligible students attended  $\geq$1 session&0.000&0.01&0.004&0.011&0.016&4466 \tabularnewline
Sessions attended by ineligible students&0.000&0.089&0.038&0.022&0.028&4466 \tabularnewline
\midrule\multicolumn{7}{c}{Panel B: parents' workshops}\tabularnewline\midrule
Eligible parents attended $\geq$1 ses.&0.048&0.487&0.039&0.000&0.000&1238 \tabularnewline
Sessions attended by eligible parents&0.099&0.933&0.107&0.000&0.000&1238 \tabularnewline
Ineligible parents attended $\geq$1 ses.&0.000&0.008&0.004&0.039&0.043&4466 \tabularnewline
Sessions attended by ineligible parents&0.000&0.016&0.008&0.062&0.062&4466 \tabularnewline
\bottomrule

\end{tabular}
  \begin{tablenotes}[para,flushleft]
\item
\footnotesize {\it Notes:} This table reports results from OLS regressions of several dependent variables on a treatment indicator. For student-level dependent variables, the regression includes lottery fixed effects. For class-level dependent variables, the regression is computed with propensity score weights. Column (1) reports the mean of the outcome variable for the control group. Column (2) reports the coefficient of the treatment indicator. Column (3) reports the standard error of this coefficient, clustered at the lottery level. Column (4) reports the unadjusted p-value of this coefficient, while Column (5) reports its p-value adjusted for multiple testing, following the method proposed in \cite{benjamini1995controlling}. Finally, Column (6) reports the number of observations used in the regression. All the dependent variables come from JUNAEB's program implementation data sets.
\end{tablenotes}
}
  \end{threeparttable}
\end{table}

\subsection{Internal validity}
\label{subsec:Internal_validity}

\subsubsection*{Balancing checks}

We test for baseline differences between the treatment and control groups by estimating Regression \eqref{eq:regression1} with student- and teacher-level baseline measures as the dependent variables. First, Table \ref{tab:H4} compares eligible students in the treatment and control groups on 29 baseline characteristics. Only two differences are significant at the 10\% level: treatment group students are more disruptive as per enumerators' ratings, and they are more likely not to be nominated as a friend by any other student in the class. Only the first of those two differences is significant at the 5\% level. 
Second, Table \ref{tab:H7} compares ineligible students in the treatment and control groups on the same 29 baseline characteristics. Four differences are significant at the 10\% level, one of which is also significant at the 5\% level. Treatment group students have slightly worse social contact, attention and focus, activity level, and disruptiveness TOCA scores.
Table \ref{tab:H10} compares teachers in the treatment and control groups on 12 characteristics.
Only one difference is significant at the 5\% level.
Finally, Table \ref{tab:H12} compares six class-level characteristics in the treatment and control groups. Three differences are significant at the 10\% level, one of which is significant at the 5\% level. Treated classes are more disruptive than control ones according to teachers and enumerators, and have higher decibel levels.

\medskip
Overall, we conduct 76 balancing checks in Tables \ref{tab:H4}, \ref{tab:H7}, \ref{tab:H10}, and \ref{tab:H12}. We find 10 significant differences between the treatment and control groups at the 10\% level, four significant differences at the 5\% level, and no significant difference at the 1\% level.

\subsubsection*{Attrition}

In this section, we document the percentage of students in our sample for which endline measures are not available, and the most common reasons for such attrition. We also show that the treatment and the control groups do not present differential levels of attrition, and that the characteristics of treatment and control group students for which endline measures are available are still balanced.

\medskip
Table \ref{tab:H1} considers attrition among eligible students.
Column (1) shows the levels of attrition in the control group. Endline measures collected by the enumerators are missing for 25.2\% of students. For 5.9\% of them this is because they have left the class between baseline and endline, for instance because their parents have moved to a different neighborhood. For the most part, the remaining 19.3\% are students who were absent on the day when the enumerators visited the class.\footnote{There are also a couple of classes that enumerators could not visit at endline, because the school principal did not want to sacrifice again a half day of instruction for the purpose of the study.}
The teacher's endline disruptiveness rating is missing for 23.2\% of students. Again, for some of them this is because they have left the class at endline. But for the majority of students, this is because their teachers refused to rate students' disruptiveness, or only rated, say, the first half of the class and then stopped because they thought the task was too time-consuming.
Column (2) of Table \ref{tab:H1} shows tests of differential attrition between the treatment and control groups, conducted by estimating Regression \eqref{eq:regression1} with measures of attrition as the dependent variables. Attrition does not seem differential: of the five measures we consider, only one is significantly different between the treatment and control groups at the 10\% level.

\medskip
Table \ref{tab:H2} considers attrition among ineligible students.
Columns (1) and (2) respectively show the levels of attrition in the control group, as well as tests for differential attrition between the treatment and control groups. The attrition levels in the control group are similar to those observed among eligible students. Here again, attrition is not differential: of the five measures we consider, only one is significantly different between the treatment and control groups at the 10\% level.

\medskip
Finally, we conduct balancing checks again, among the students whose endline measures are available. Table \ref{tab:H5} (resp. Table \ref{tab:H6}) considers the same 29 baseline characteristics as in Table \ref{tab:H4}, and compares their mean in the treatment and control groups, among the eligible students for which enumerators' endline measures (resp. the teacher's endline disruptiveness rating) are (resp. is) available. As in Table \ref{tab:H4}, few differences are significant. Table \ref{tab:H8} repeats the same exercise, among ineligible students for which enumerators' endline measures are available. Again, few differences are significant. Finally, Table \ref{tab:H9} compares ineligible students for which the teacher's endline disruptiveness rating is available in the treatment and control groups. More differences are significant, but most become insignificant once p-values are adjusted for multiple testing. Overall, the post-attrition treatment and control group students whose outcomes are compared in Section \ref{sec:Treatment Effects} seem to have balanced baseline characteristics.

\medskip
Turning to class-level attrition, while we have teachers' and enumerators' ratings of classes' disruptiveness for more than 90\% of classes in our sample, we have some differential attrition for teachers' questionnaires: none is missing in the control group, while 8\% are missing in the treatment group, and the difference is statistically significant. In Table \ref{tab:H12_03}, we conduct again the balancing checks on the baseline class-level measures in Table \ref{tab:H12}.\footnote{Table \ref{tab:H12_03} was not pre-specified in our PAP, because we had not anticipated the possibility of differential attrition for the class-level measures.} For measures made by teachers, we restrict the sample to classes for which all class-level endline teacher measures are available, while for measures made by enumerators we restrict the sample to classes for which all class-level endline enumerators measures are available. As in Table \ref{tab:H12}, three differences are significant at the 10\% level, but none is significant at the 5\% level.

\subsection{Estimation methods}
\label{subsec:Estimationmethods}

In this section, we discuss the methods we use to estimate the effect of the treatment. For our student-level outcomes, we estimate the following regression:
\begin{equation}\label{eq:regression2}
Y_{ijk}=\gamma_k+X'_{ijk}\theta_1+\beta D_{jk}+u_{ijk},
\end{equation}
where $Y_{ijk}$ is the outcome of student $i$ in class $j$ and lottery $k$, the $\gamma_k$s are lottery fixed effects, $X_{ijk}$ denotes student-level baseline variables used as statistical controls, and $D_{jk}$ is an indicator variable equal to 1 if class $j$ in lottery $k$ was assigned to the treatment group. $\widehat{\beta}$ estimates the ITT effect of being assigned to the treatment on the outcome. As in Regression \eqref{eq:regression1}, we cluster the standard errors at the lottery level. To select the controls, we follow \cite{belloni2014high}. We run a Lasso regression of the outcome on all the student-level baseline variables in Table \ref{tab:H4}, and we pick the variables selected by the Lasso.\footnote{In a randomized experiment, the treatment is by construction uncorrelated with the controls, so it is not necessary to run a Lasso regression of the treatment on the controls.}

\medskip
For all the class-level outcomes, we estimate the following regression:
\begin{equation}\label{eq:regression3}
Y_{jk}=\alpha+Z'_{jk}\theta+\beta D_{jk}+u_{jk},
\end{equation}
where $Y_{jk}$ is the outcome of class $j$ in lottery $k$, $Z_{jk}$ denotes class-level baseline variables used as statistical controls, and $D_{jk}$ is the treatment indicator. The regression is weighted by propensity score weights, and as in Regression \eqref{eq:regression1}, we cluster the standard errors at the lottery level. To select the controls, we follow again \cite{belloni2014high}, and we run a Lasso regression of the outcome on the class average of all the student-level baseline variables in Table \ref{tab:H4}, and all the class-level baseline variables in Tables \ref{tab:H10} and \ref{tab:H12}, and we pick the variables selected by the Lasso.

\medskip
To account for multiple testing, we follow the same approach as \cite{finkelstein2010}. First, we group related outcomes into hypothesis. For instance, students' happiness, self-esteem, and self-control scores are grouped together into an ``emotional stability'' hypothesis. Then, for each outcome, we report both the unadjusted p-value of the estimated effect, and the adjusted p-value controlling the FDR within the hypothesis the outcome belongs to. Each panel in Tables \ref{tab:H14}, \ref{tab:H21}, and \ref{tab:H25}
corresponds to a set of related outcomes grouped into an hypothesis. Finally, for each hypothesis we also report the effect of the treatment on a weighted average of the outcomes in that hypothesis, using the weights proposed in \cite{anderson2008multiple}. We refer to the effect of the treatment on this weighted average as the standardized treatment effect.

\section{Treatment Effects}
\label{sec:Treatment Effects}

\subsection{Effects on eligible students}
\label{subsec:effectdisruptive}

Table \ref{tab:H14} below shows the effect of the SFL workshops on eligible students' outcomes.

\begin{table}[htbp] \centering
  \begin{threeparttable}
  \caption{Treatment effect on eligible students}
\label{tab:H14}
{\normalsize
\begin{tabular}{lccccccc}
\toprule\toprule
{Variables}&{Control}&{T-C}&{S.E.}&{Unadj. P}&{Adj. P}&{N} \tabularnewline
&{(1)}&{(2)}&{(3)}&{(4)}&{(5)}&{(6)} \tabularnewline
\midrule\multicolumn{7}{c}{Panel A: emotional stability}\tabularnewline\midrule
School happiness score&-0.107&0.123&0.075&0.101&0.304&876 \tabularnewline
Self-control score&-0.184&-0.04&0.087&0.648&0.648&880 \tabularnewline
Self-esteem score&-0.17&-0.106&0.079&0.176&0.264&903 \tabularnewline
Standardized Treatment Effect&0.015&-0.002&0.08&0.977&&915 \tabularnewline
\midrule\multicolumn{7}{c}{Panel B: disruptiveness}\tabularnewline\midrule
Disruptiveness, teacher&0.353&0.1&0.102&0.327&0.654&904 \tabularnewline
Disruptiveness, enumerator&0.157&0.02&0.083&0.805&0.805&948 \tabularnewline
Standardized Treatment Effect&-0.025&0.062&0.089&0.489&&1110 \tabularnewline
\midrule\multicolumn{7}{c}{Panel C: academic outcomes}\tabularnewline\midrule
\% school days missed&12.82&1.055&1.016&0.299&0.896&1236 \tabularnewline
Spanish test score&-0.308&-0.049&0.069&0.482&0.723&956 \tabularnewline
Math test score&-0.274&-0.006&0.08&0.945&0.945&956 \tabularnewline
Standardized Treatment Effect&0.011&-0.035&0.071&0.622&&1238 \tabularnewline

\midrule\multicolumn{7}{c}{Panel D: integration in the class network}\tabularnewline\midrule
No friends in the class&0.27&-0.028&0.027&0.307&0.409&1147 \tabularnewline
\% class friends with student&0.07&0.007&0.005&0.145&0.291&1147 \tabularnewline
Friends' average ability&-0.061&-0.011&0.077&0.883&0.883&829 \tabularnewline
Friends' average disruptiveness&0.177&0.132&0.087&0.131&0.525&787 \tabularnewline

Standardized Treatment Effect&-0.008&0.038&0.063&0.54&&1148 \tabularnewline

\bottomrule
\end{tabular}
  \begin{tablenotes}[para,flushleft]
\footnotesize {\it Notes:} This table reports results from OLS regressions of several dependent variables on a treatment indicator, lottery fixed effects, and control variables for eligible students. The control variables are selected by a Lasso regression of the dependent variable on all potential controls, following \cite{belloni2014high}. Column (1) reports the mean of the outcome variable for the control group. Column (2) reports the coefficient of the treatment indicator. Column (3) reports the standard error of this coefficient, clustered at the lottery level. Column (4) reports the unadjusted p-value of this coefficient, while Column (5) reports its p-value adjusted for multiple testing, following the method proposed in \cite{benjamini1995controlling}. Finally, Column (6) reports the number of observations used in the regression. All the dependent variables, except for {\it \% school days missed}, were collected by the authors at endline.
\end{tablenotes}
}
  \end{threeparttable}
\end{table}

Panel A shows that the SFL workshops do not have large effects on eligible students' emotional stability.
The average school happiness score is $0.123\sigma$ higher in the treatment than in the control group, but this difference is not very significant (p-value=0.101), and becomes insignificant after adjusting for multiple testing. The average self-esteem score is $0.106\sigma$ lower in the treatment group, but this difference is insignificant even before adjusting for multiple testing (p-value=0.176).
The average self-control score is very close in the treatment and control groups. Finally, the average standardized score is also very close in the treatment and control groups.

\medskip
Panel B shows that SFL does not have a large effect on eligible students' disruptiveness. At endline, the average teachers' disruptiveness rating is $0.1\sigma$ higher in the treatment than in the control group. This difference is not statistically significant at conventional levels, but based on its estimated standard error, we can rule out at the 5\% level that SFL reduces teachers' disruptiveness ratings by more than $0.1\sigma$. This is around 1/5 of the treatment effect on students' disruptiveness found by \cite{payton2008} in their meta-analysis of selected SEL programs. Enumerators' disruptiveness ratings also do not significantly differ in the treatment and control groups.


\medskip
Panel C shows that SFL also does not have large effects on the academic outcomes of eligible students. For instance, students' Spanish and mathematics scores are very close in the two groups. We can reject at the 5\% level that SFL increases eligible students' Spanish and mathematics scores by more than $0.086\sigma$ and $0.151\sigma$, respectively. Again, these effects are much smaller than those found in the meta-analysis by \cite{payton2008}.

\medskip
Finally, Panel D shows that SFL does not have large effects on eligible students' friendship ties. The proportion of students not nominated as a friend by any other student in the class is 2.8 percentage points lower in the treatment than in the control group, but this difference is insignificant.

\medskip
Overall, we do not find evidence of a positive effect of SFL on any of the dimensions we consider, and we can also rule out much smaller effects than those previously found for similar programs.

\subsection{Effects on ineligible students}
\label{subsec:effectclassmates}

In this section, we explore whether the SFL workshops have spillover effects on ineligible students. Panel A of Table \ref{tab:H21} below shows that these workshops do not generate strong spillover effects on the emotional stability of ineligible students. The average school happiness, self-control, and self-esteem scores are very close and do not significantly differ in the treatment and control groups.

\begin{table}[H] \centering
  \begin{threeparttable}
\caption{Treatment effect on ineligible students}
\label{tab:H21}
{\normalsize
\begin{tabular}{lccccccc}
\toprule\toprule
{Variables}&{Control}&{T-C}&{S.E.}&{Unadj. P}&{Adj. P}&{N} \tabularnewline
&{(1)}&{(2)}&{(3)}&{(4)}&{(5)}&{(6)} \tabularnewline
\midrule\multicolumn{7}{c}{Panel A: emotional stability}\tabularnewline\midrule
School happiness score&0.026&0.016&0.037&0.666&0.666&3360 \tabularnewline
Self-control score&0.097&-0.05&0.043&0.25&0.751&3404 \tabularnewline
Self-esteem score&0.084&-0.043&0.047&0.36&0.54&3446 \tabularnewline
Standardized Treatment Effect&0.027&-0.023&0.042&0.577&&3476 \tabularnewline

\midrule\multicolumn{7}{c}{Panel B: disruptiveness}\tabularnewline\midrule
Disruptiveness, teacher&-0.212&0.208&0.106&0.05&0.101&3203 \tabularnewline
Disruptiveness, enumerator&-0.046&-0.003&0.046&0.954&0.954&3518 \tabularnewline
Standardized Treatment Effect&-0.051&0.063&0.072&0.384&&4033 \tabularnewline

\midrule\multicolumn{7}{c}{Panel C: academic outcomes}\tabularnewline\midrule
\% school days missed&13.089&0.382&0.634&0.547&0.82&4427 \tabularnewline
Spanish test score&0.128&-0.055&0.055&0.316&0.948&3517 \tabularnewline
Math test score&0.08&-0.013&0.056&0.821&0.821&3517 \tabularnewline
Standardized Treatment Effect&0.018&-0.019&0.044&0.66&&4452 \tabularnewline
\midrule\multicolumn{7}{c}{Panel D: integration in the class network}\tabularnewline\midrule
No friends in the class&0.197&-0.035&0.013&0.008&0.033&4168 \tabularnewline
\% class friends with student&0.087&0.004&0.003&0.156&0.312&4168 \tabularnewline
Friends' average ability&0.027&-0.011&0.077&0.884&0.884&3342 \tabularnewline
Friends' average disruptiveness&-0.11&0.051&0.053&0.338&0.45&3176 \tabularnewline
Standardized Treatment Effect&0.003&0.066&0.037&0.076&&4171 \tabularnewline

\bottomrule
\end{tabular}
  \begin{tablenotes}[para,flushleft]
\footnotesize {\it Notes:} This table reports results from OLS regressions of several dependent variables on a treatment indicator, lottery fixed effects, and control variables for ineligible students. The control variables are selected by a Lasso regression of the dependent variable on potential controls, following \cite{belloni2014high}. Column (1) reports the mean of the outcome variable for the control group. Column (2) reports the coefficient of the treatment indicator. Column (3) reports the standard error of this coefficient, clustered at the lottery level. Column (4) reports the unadjusted p-value of this coefficient, while Column (5) reports its p-value adjusted for multiple testing, following the method proposed in \cite{benjamini1995controlling}. Finally, Column (6) reports the number of observations used in the regression. All the dependent variables, except for {\it \% school days missed}, were collected by the authors at endline.
\end{tablenotes}
}
  \end{threeparttable}
\end{table}

Panel B suggests that the SFL workshops may generate negative spillover effects on ineligible students' disruptiveness. At endline, the average of teachers' disruptiveness ratings is $0.208\sigma$ higher in the treatment than in the control group. This difference is significant (p-value=0.05), but becomes marginally insignificant after adjusting for multiple testing (adjusted p-value=0.101).

\medskip
Then, Panel C shows that the SFL workshops do not have large spillover effects on ineligible students' academic outcomes. 
Finally, Panel D shows that SFL improves the integration of ineligible students in the class network. The proportion of students not nominated as a friend by any other student in the class is 3.5 percentage points lower in the treatment than in the control group, a 17.8\% reduction in the fraction of ineligible students who have no friends. This difference is significant (p-value=0.008), and it remains significant after accounting for multiple testing (adjusted p-value=0.033). Similarly, ineligible students are nominated as friends by 9.1\% of their classmates in the treatment group, against 8.7\% in the control group, but this difference is not significant. The treatment does not significantly alter the academic ability and disruptiveness of ineligible students' friends. Finally, the average standardized score constructed from these four outcomes is significantly higher in the treatment than in the control group (p-value=0.076).

\subsection{Effects on the classroom environment}
\label{subsec:effectclassroom}

\begin{table}[htbp] \centering
  \begin{threeparttable}
\caption{Treatment effect on classroom environment}
\label{tab:H25}
{\normalsize
\begin{tabular}{lccccccc}
\toprule\toprule
{Variables}&{Control}&{T-C}&{S.E.}&{Unadj. P}&{Adj. P}&{N} \tabularnewline
&{(1)}&{(2)}&{(3)}&{(4)}&{(5)}&{(6)} \tabularnewline\midrule
Disruptiveness, teacher&-0.187&0.232&0.137&0.091&0.226&160 \tabularnewline
Bullying in class, teacher&-0.038&0.105&0.153&0.492&0.492&160 \tabularnewline
Disruptiveness, enumerator&-0.186&0.389&0.148&0.009&0.043&167 \tabularnewline
Delay in class's start (minutes)&9.938&1.204&1.046&0.25&0.312&160 \tabularnewline
Average decibels during class&0.022&0.681&0.487&0.162&0.27&169 \tabularnewline
Standardized Treatment Effect&-0.215&0.424&0.131&0.001&&169 \tabularnewline
\bottomrule
\end{tabular}

  \begin{tablenotes}[para,flushleft]
\item \footnotesize {\it Notes:} This table reports results from OLS regressions of several dependent variables on a treatment indicator and control variables, computed with propensity score weights. The control variables are selected by a Lasso regression of the dependent variable on all potential controls, following \cite{belloni2014high}. Column (1) reports the mean of the outcome variable for the control group. Column (2) reports the coefficient of the treatment indicator. Column (3) reports the standard error of this coefficient, clustered at the lottery level. Column (4) reports the unadjusted p-value of this coefficient, while Column (5) reports its p-value adjusted for multiple testing, following the method proposed in \cite{benjamini1995controlling}. Finally, Column (6) reports the number of observations used in the regression. All the dependent variables were collected by the authors at endline.
\end{tablenotes}
}
  \end{threeparttable}
\end{table}
In this section, we study how the SFL workshops affect different measures of classrooms' environment at endline. Table \ref{tab:H25} above shows that SFL worsens teachers' and enumerators' disruptiveness ratings of the classes. Those ratings are based on teachers' and enumerators' agreement with statements like ``There is noise and disorder in this class,'' or ``The teacher has to wait for a long time before students calm down and he/she can start teaching''. According to teachers, treated classes are $0.232\sigma$ more disruptive than control ones. This difference is statistically significant before adjusting for multiple testing (p-value=0.091), but it becomes insignificant after adjusting for it (adjusted p-value=0.226). According to enumerators, treated classes are $0.389\sigma$ more disruptive. This difference is statistically significant before and after adjusting for multiple testing (p-value=0.009, adjusted p-value=0.043).
Enumerators do not know if the class they observe has been treated or not, contrary to teachers. The fact that they also find that treated classes are more disruptive suggests that teachers' worse perception of the treatment-group classes is not a mere placebo effect. Table \ref{tab:H12} in the Appendix shows that treated and control classes are imbalanced on these two measures at baseline, so we reestimate these two effects controlling for these two measures.\footnote{In the estimation of the treatment effect on teachers' ratings, the Lasso selects teachers' baseline ratings as a control, but it does not select enumerators' ratings. In the estimation of the treatment effect on enumerators' ratings, the Lasso does not select any control.} The estimated treatment effects on teachers' and enumerators' ratings are now respectively equal to $0.247\sigma$ (p-value=0.084) and $0.282\sigma$ (p-value=0.066), so the treatment effects on these two measures do not seem due to imbalances already existing at baseline.

\medskip
It may be surprising that the treatment significantly worsens enumerators' ratings of classes' overall disruptiveness, without affecting their ratings of eligible and ineligible students' disruptiveness, as shown in Panel B of Tables \ref{tab:H14} and \ref{tab:H21}. While the limited amount of time they spend in each classroom may be enough for them to observe that there is more disorder in the treated classes, it may not be sufficient for them to pinpoint the students responsible for that disorder.

\medskip
Table \ref{tab:H25} also shows that treated classes have higher levels of bullying, that their lectures start 1.2 more minutes after the scheduled time than in control classes, and that they have higher levels of decibels. Even though these results are not statistically significant, they go in the same direction as the results on the disruptiveness measures.

\medskip
Finally, the average standardized score constructed from the five outcomes in Table \ref{tab:H25} is $0.424\sigma$ higher in the treatment than in the control group. This difference is highly significant (p-value=0.001), and it remains highly significant even accounting for the fact that in Tables \ref{tab:H14}, \ref{tab:H21}, and \ref{tab:H25} we estimate the effect of the treatment on nine standardized scores (adjusted p-value=0.009). Therefore, we can conclude that SFL significantly worsens the studying conditions in treated classes.

\subsection{Robustness checks}
\label{subsec:heterogeneouseffects}

As a robustness check, we reestimate all the regressions in Tables \ref{tab:H14}, \ref{tab:H21}, and \ref{tab:H25}
without controls. The results of that exercise can be found in Tables \ref{tab:H14X}, \ref{tab:H21X}, and \ref{tab:H25X}.
Results with and without controls are pretty similar, except that the effects on ineligible students' friendships are no longer significant without controls. In our PAP, we had indicated that as a further robustness check, we would recompute all the unadjusted p-values in Tables \ref{tab:H14}, \ref{tab:H21}, and \ref{tab:H25}
using randomization inference. Doing so does not change our main findings so the results of that exercise are not reported here but are available upon request.

\section{Interpretation and exploratory analysis}
\label{sec:mech}

Table \ref{tab:H14} shows that SFL does not have positive effects on eligible students' emotional stability, disruptiveness, and academic ability. This is at odds with an extensive literature, that has shown that selected SEL programs usually produce large positive effects on these dimensions. In a meta-analysis of 80 selected SEL interventions, \cite{payton2008} find that they reduce conduct problems  by $0.47\sigma$, and respectively  improve emotional stability and academic performance by $0.50$ and $0.43\sigma$.
Based on our estimates, we can reject effects much smaller than those found in \cite{payton2008}.

\medskip
\textnormal{
There are several other meta-analyses of SEL interventions that have been peer-reviewed, unlike \cite{payton2008}, and that are more recent. However, they either focus on universal interventions delivered to the whole class rather than to a selected group of students (see \citep{durlak2011impact}, \citep{sklad2012effectiveness}, \citep{wigelsworth2016impact}, \citep{taylor2017promoting}, and \citep{corcoran2018effective}), or they include both universal and selected interventions but do not report effects separately for both types of interventions (see \citep{dymnicki2012adolescent}). To our knowledge, \cite{payton2008} is the only meta-analysis reporting effects separately for selected SEL interventions comparable to SFL, which is why we focus on that meta-analysis.
In any case, those six other meta-analyses also find pretty large effects, even though they are slightly lower than those in \cite{payton2008}. The effects they find on conduct problems range from -0.14 to -0.47$\sigma$, with an average equal to -0.25$\sigma$. Similarly, effects on emotional stability range from 0.23 to 0.74$\sigma$, and the average is 0.55$\sigma$. Finally, effects on academic performance range from 0.26 to 0.53$\sigma$, and the average is 0.28$\sigma$. Therefore, we can still reject effects substantially smaller than the average effects in those meta-analyses. Overall, our results are at odds with a very substantial literature that has studied SEL programs.}

\medskip
\textnormal{
To understand this discrepancy,
in Table \ref{tab:metanalysis} below we compare SFL to the selected SEL interventions reviewed in \cite{payton2008}, to assess if SFL differs from those interventions in any striking way that could account for its lower effect.} Many features of the interventions reviewed in \cite{payton2008} are readily available from Table 7 therein. Other features that seemed important to us are not reported in the paper, so we reviewed a random sample of 25 of the meta-analysis's papers, and manually collected those features. They appear in italic in Table \ref{tab:metanalysis}.

\begin{table}[htbp] \centering
  \begin{threeparttable}
  \caption{Comparing ``Skills for Life'' to the selected SEL interventions in \cite{payton2008}}
\label{tab:metanalysis}
{\normalsize
\begin{tabular}{lcc}
\toprule\toprule
{}&{Skills for Life}&{\cite{payton2008}}\tabularnewline
\midrule\multicolumn{3}{c}{Panel A: Intervention Intensity}\tabularnewline\midrule
\textit{Number of sessions} & 10 & 12 (median) \\
\textit{Sessions' duration in minutes} & 120 & 50  (median) \\
\textit{Intervention duration in weeks} & 10 & 10 (median) \\
\textit{Number of students per workshop} & 7.2 & 6 (median) \\
Parental training & Yes & 41\% \\
\textit{Parental sessions} & 3 & 14 (median) \\
\textit{Parents' attendance} & 34\% & 49\% (median) \\
Students pulled out of class & Yes & 100\% \\
\midrule\multicolumn{3}{c}{Panel B: Targeting of eligible students}\tabularnewline\midrule
Primary school students & Yes & 69\% \\
Students with conduct problems & Yes & 48\% \\
Students with emotional problems & Yes & 23\% \\
Students with conduct and emotional problems & Yes & 29\% \\
\textit{Low SES students} & Yes & 73\% \\
\midrule\multicolumn{3}{c}{Panel C: Study design}\tabularnewline\midrule
Random assignment of treatment & Yes & 80\% \\
\textit{Journal's impact factor (for published studies)} & NA & 4.01 (median) \\
Outcomes based on teacher ratings per study & 3 & 0.6 \\
Outcomes based on enumerator ratings per study & 2 & 0.3 \\
Outcomes based on student ratings per study & 5 & 1.2875  \\
Outcomes based on parent ratings per study & 0 & 0.175 \\
Outcomes based on school records per study & 1 & 0.35 \\
Uses validated psychometric scale as outcome & Yes & 69\% \\
\textit{Weeks between end of intervention and endline} & 3 & 1 (median) \\
\midrule\multicolumn{3}{c}{Panel D: Location}\tabularnewline\midrule
United States & No & 85\% \\
\textit{High-income country} & No & 100\% \\
\midrule\multicolumn{3}{c}{Panel E: Delivery Personnel, Monitoring of Delivery, and Intervention Scale}\tabularnewline\midrule
Intervention delivered by: &  & \\
\textit{Researchers (alone, or together with school staff)} & No & 43\% \\
\textit{School staff trained and monitored by researchers} & No & 22\% \\
\textit{Other personnel trained and monitored by researchers} & No & 35\% \\
\textit{Frequency at which delivery is monitored:} & Never & Weekly (median)
\\
\textit{Number of treated students} & 8,570 & 36 (median) \\
\bottomrule
\end{tabular}
  \begin{tablenotes}[para,flushleft]
\footnotesize {\it Notes:} This table compares the ``Skills for Life'' intervention to those in the meta-analysis of \cite{payton2008}. For the meta-analysis's papers, the variables in italic were collected manually by the authors, by reviewing a random sample of 25 of the 80 articles reviewed by \cite{payton2008}. The variables not in italic are directly available from Table 7 in \cite{payton2008}. SFL's number of treated students is for 2013.
\end{tablenotes}
}
  \end{threeparttable}
\end{table}

\subsection{SFL's intensity is comparable to that of meta-analysis's interventions}

Panel A of Table \ref{tab:metanalysis} shows that SFL's intensity is similar to that of the meta-analysis's interventions. The median number of sessions across those interventions is slightly higher than SFL's number of sessions (12 versus 10), but their sessions are typically shorter (50 versus 120 minutes). The number of students per workshop is comparable (a median of 6 in the meta-analysis, versus 7.2 on average in our sample). Their median duration is the same as SFL's (10 weeks). 59\% of those interventions only include sessions with students, while 41\% also include a parental training, like SFL. Only seven of the papers we reviewed give the number of parental sessions, but among those the median number of sessions (14) is higher than in SFL (three parental sessions). Only three of the papers we reviewed mention parents' attendance, but among those the median attendance (49\%) is comparable to that in SFL (34\%, see Table \ref{tab:H13}). Note also that \cite{payton2008} do not mention that the presence and intensity of a parental training is correlated with larger program effects. In all those interventions, selected students are pulled-out of their class during the class day, as in the SFL intervention.\footnote{The three SFL teams we interviewed said that schools' cooperation is usually very good, and that they do not have issues scheduling and delivering the sessions.}

\medskip
Panel B of Table \ref{tab:metanalysis} shows that SFL uses similar criteria as the programs reviewed by \cite{payton2008} to determine which students are eligible. Like SFL, 69\% of those interventions treat primary school students. 48\% target students with conduct problems, 23\% target students with emotional distress, and the remaining interventions target students with a combination of problems. 73\% target low SES students, like SFL.

\subsection{Our study design is comparable to that of metanalysis's studies}

Our study design is also comparable to that of the meta-analysis's studies. Panel C of Table \ref{tab:metanalysis} shows that the treatment was randomly assigned in 80\% of those studies. \textnormal{Many of the published studies appeared in high-impact-factor peer-reviewed journals (median impact factor=4.01).\footnote{85\% of the 80 studies reviewed by \cite{payton2008} were published in peer-reviewed journals.}} Most of their outcome measures are teacher, enumerator, and student ratings, often made using validated psychometric scales, as in our study. We measured our outcomes three weeks after the end of the SFL intervention, while in the reviewed interventions, the median number of weeks between the end of the intervention and endline data collection is equal to one.

\medskip
Another methodological concern is that our control group may have benefited from the treatment, as we have some schools that have both treated and control classes, and treated students may interact with students from control classes in their school. To assess if this is a serious concern, we estimate SFL's effect in schools where only one class was included in our experiment. In this subsample, which still has 114 classes, we find that teachers' ratings of eligible students' disruptiveness is 0.2$\sigma$ higher in the treatment than in the control group (p-value=0.12), and we can rule out at the 5\% level that SFL reduces eligible students' disruptiveness by more than 0.06$\sigma$. Results are similar when we consider other outcomes, such as students' test scores. Overall, control-group contamination seems unlikely to account for SFL's lack of effect.

\subsection{Students receiving SFL may be harder to treat than those in the meta-analysis's interventions.}

Panel D of Table \ref{tab:metanalysis} shows that 85\% of the interventions in \cite{payton2008} take place in the US, and all take place in high-income countries. SFL takes place in Chile, and may then be faced with a harder-to-treat population than those interventions. For instance, recent epidemiological studies show that the prevalence rate of ADHD, a disorder correlated with conduct problems, is equal to 15.5\% among primary school children in Chile (see \citep{de2013epidemiology}), against 6.8\% in the US (see \citep{visser2014trends}), 3.5 to 5.6\% in France (see \citep{lecendreux2011prevalence}) or 3\% in Italy (see \citep{bianchini2013prevalence}). Similarly, surveys indicate that domestic violence, a cause of conduct disorder problems in children (see \citep{carrell2010externalities}), is more prevalent in Chile than in the US. 4.3\% of Chilean women report having been physically assaulted by their partner over the previous year (see \citep{ministerio2017}), against 1.3\% in the US (see \citep{tjaden2000}). Then, disruptive students may suffer from more severe problems in Chile than in high-income countries and may be harder to treat. This could explain why SFL produces lower effects than SEL programs in high-income countries.

\medskip
To test this hypothesis, we start by assessing whether SFL's effect is stronger for less disruptive students. Specifically, we look at the effect of SFL for eligible students with a TOCA score below the median. If anything, we find slightly negative effects: the program increases their teachers' disruptiveness ratings by $0.18\sigma$, but this effect is marginally significant (p-value=0.09).

\medskip
Then, as primary school students with serious behavioral problems are more present in Chile than in high-income countries, each SFL workshop is more likely to comprise some very disruptive students than an SEL workshop in a high-income country, and the presence of those hard-to-treat students may lower the workshop's effectiveness for every student, including the less disruptive ones. To investigate that possibility, we computed the 90th percentile\footnote{The choice of the 90th percentile was guided by the fact that $0.9^7=48\%$, so assuming that students' disruptiveness levels are independent within a class and that all classes have 7 eligible students, 52\% of classes should have at least one student above that percentile. In practice, the proportion of classes that have at least one eligible student above that percentile is slightly lower (46\%), but still close to 50\%.} of the average of the authority acceptance, attention and focus, activity levels, and overall disruptiveness TOCA scores among eligible students, and estimated SFL's effects in the 79 classes that have at least one very disruptive eligible student above that threshold. Those classes have 123 very disruptive eligible students, 534 other eligible students, and  2,064 ineligible students. In Table \ref{tab:verydis} below, we estimate the effects separately for each group of students, focusing on disruptiveness ratings and test scores, and on the friendship nominations received by very disruptive eligible students. Unadjusted p-values and p-values controlling the False Discovery Rate (FDR) (see \citep{benjamini1995controlling}) across all the tests in the table are presented.

\medskip
First, Panel A of the table shows that the program does not have any statistically significant effect on the disruptiveness and test scores of very disruptive eligible students, but increases by 50\% the percentage of their classmates who nominate them as friends (unadjusted p-value=0.042, adjusted p-value=0.102). Second, in Panel B we estimate SFL's effects among the other eligible students. The program increases their teachers' disruptiveness ratings by 0.496$\sigma$ (unadjusted p-value=0.0006, adjusted p-value=0.0102), may reduce their Spanish scores by 0.201$\sigma$ (unadjusted p-value=0.033, adjusted p-value=0.112), does not have a significant effect on their enumerators' disruptiveness ratings and maths scores, and doubles the proportion that nominate at least one very disruptive student as a friend (unadjusted p-value=0.012, adjusted p-value=0.051). Very disruptive eligible students may then have a negative influence on other eligible students, which could explain the negative effects the program has on them. Third, in Panel C we estimate that the program increases teachers' and enumerators' disruptiveness ratings of ineligible students, respectively by 0.477$\sigma$ (unadjusted p-value=0.008, adjusted p-value=0.045) and 0.137$\sigma$ (unadjusted p-value=0.083, adjusted p-value=0.176). On the other hand, the program does not have a significant effect on the test scores of those students and on the proportion of them who nominate a very disruptive student as a friend. The mechanism whereby the program makes ineligible students more disruptive may be a contagion effect: eligible students become more disruptive, and ineligible students imitate them. Finally, in Panel D we estimate that the program increases teachers' and enumerators' overall disruptiveness ratings of the classes, respectively by 0.669$\sigma$ (unadjusted p-value=0.005, adjusted p-value=0.043) and 0.516$\sigma$ (unadjusted p-value=0.035, adjusted p-value=0.099). The regressions in the table are estimated with the controls selected by the Lasso. Treatment effects are similar when those controls are dropped (see Appendix Table \ref{tab:verydis_notctrl}), and when the few covariates that are unabalanced at baseline in the relevant subsample are added as controls (see Appendix Table \ref{tab:verydis_forcedctrl}).

\medskip
Classes with at least one very disruptive eligible student have slightly more eligible students than classes that do not have any (8.3 versus 6.2). This difference is not very large, but we still checked if we also find negative effects of the program in the subsample of classes that have more eligible students than the median. The answer is negative, so it does not seem that the negative effects we find in classes with at least one very disruptive eligible student are mediated by the slightly higher number of eligible students in those classes.

\medskip
Overall, we find suggestive evidence that SFL's effectiveness is hampered by the presence of very disruptive students, who may be less present in the other contexts where SEL programs have been shown to work. We still do not find statistically significant effects of SFL in the 93 classes that do not have any very disruptive student. This may be because the effects we can reject in this subsample are too large, though we can for instance still reject at the 5\% level an effect larger than 0.13$\sigma$ on Spanish test scores. Another potential explanation is that even those classes may still have some students that are more disruptive than the typical students benefiting from selected SEL programs in the US.

\begin{table}[H] \centering
  \begin{threeparttable}
\caption{Treatment effect in classes with at least one very disruptive students}
\label{tab:verydis}
{\normalsize
\begin{tabular}{lcccccc}
\toprule\toprule
{Variables}&{Control}&{T-C}&{S.E.}&{Unadj. P}&{Adj. P}&{N} \tabularnewline
&{(1)}&{(2)}&{(3)}&{(4)}&{(5)}&{(6)} \tabularnewline
\midrule\multicolumn{6}{c}{Panel A: Very disruptive eligible students}\tabularnewline\midrule
Disruptiveness, teacher&0.985&-0.175&0.330&0.600&0.850&86\tabularnewline
Disruptiveness, enumerator&0.286&0.613&0.439&0.162&0.306&85 \tabularnewline
Spanish test score&-0.460&0.047&0.304&0.878&1&88 \tabularnewline
Math test score&-0.230&0.063&0.512&0.902&1&88 \tabularnewline
\% class friends with student &0.051&0.025&0.012&0.042&0.102&109 \tabularnewline
\midrule\multicolumn{6}{c}{Panel B: Not very disruptive eligible students}\tabularnewline\midrule
Disruptiveness, teacher&0.294&0.496&0.145&0.001&0.010&391\tabularnewline
Disruptiveness, enumerator&0.162&0.118&0.124&0.339&0.576&393 \tabularnewline
Spanish test score&-0.349&-0.201&0.097&0.033&0.112&397 \tabularnewline
Math test score&-0.349&-0.008&0.181&0.965&0.965&397 \tabularnewline
Friends with $\geq1$ very dis.& 0.065&0.075&0.030&0.012&0.051&397 \tabularnewline
\midrule\multicolumn{6}{c}{Panel C: Ineligible students}\tabularnewline\midrule
Disruptiveness, teacher&-0.205&0.477&0.181&0.008&0.045&1517\tabularnewline
Disruptiveness, enumerator&-0.093&0.137&0.079&0.083&0.176&1576 \tabularnewline
Spanish test score&0.035&0.012&0.122&0.924&0.982&1579 \tabularnewline
Math test score&0.115&0.053&0.151&0.725&0.948&1579 \tabularnewline
Friends with $\geq1$ very dis.& 0.067&0.015&0.028&0.584&0.903&1577 \tabularnewline
\midrule\multicolumn{6}{c}{Panel D: Class-level outcomes}\tabularnewline\midrule
Disruptiveness, teacher&-0.250&0.669&0.236&0.005&0.043&72\tabularnewline
Disruptiveness, enumerator&-0.250&0.516&0.245&0.035&0.099&76 \tabularnewline
\bottomrule
\end{tabular}
  \begin{tablenotes}[para,flushleft]
\footnotesize {\it Notes:} This table reports results from OLS regressions of several dependent variables on a treatment indicator and control variables. The control variables are selected by a Lasso regression of the dependent variable on potential controls, following \cite{belloni2014high}. To account for the fact the randomization is stratified, the regressions in Panels A, B, and C have lottery fixed effects, while in the regressions in Panel D we use propensity score reweighting. Column (1) reports the mean of the outcome variable for the control group. Column (2) reports the coefficient of the treatment indicator. Column (3) reports the standard error of this coefficient, clustered at the lottery level. Column (4) reports the unadjusted p-value of this coefficient. Finally, Column (5) reports the number of observations used in the regression. All the dependent variables were collected by the authors at endline.
\end{tablenotes}
}
  \end{threeparttable}
\end{table}

\subsection{SFL's delivery is less monitored than the meta-analysis interventions'.}

\medskip
\textnormal{
Panel E of Table \ref{tab:metanalysis} shows that SFL strikingly differs from the meta-analysis's programs in terms of delivery. All of the meta-analysis's interventions are demonstration programs, mounted by researchers for research purposes. 43\% of the interventions are entirely or partly delivered by the researchers, 22\% are delivered by school staff trained and supervised by the researchers, and 35\% are delivered by other personnel (most often psychologists) hired, trained, and supervised by the researchers. 69\% of the studies where the intervention was not entirely delivered by the researchers mention the frequency at which the researchers monitored the delivery personnel, for instance by attending sessions, or by reviewing video- or audio-recorded sessions. The median is a weekly monitoring. Researchers' involvement is very high in the studies reviewed by \cite{payton2008}, but another meta-analysis of universal SEL interventions suggests that they can produce large effects without researchers' involvement. \cite{wigelsworth2016impact} review 25 interventions implemented without researchers' involvement and find large effects: -0.15$\sigma$ for conduct problems, +0.47$\sigma$ for emotional stability, and +0.22$\sigma$ for academic performance. However, looking at a random sample of 10 of those 25 studies, it appears that in 6 of the 8 studies where monitoring was discussed, monitoring was frequent and intensive, and was often conducted by an NGO promoting the program. Overall, in the majority of the SEL interventions considered in those meta-analysis, delivery is monitored frequently by a third party.}

\medskip
\textnormal{
JUNAEB provides SFL implementers with a detailed manual describing the content of each of the workshop's session, and the municipal teams we interviewed said they follow this manual. SFL employees also attend ``good practices'' meetings every six months, during which they share with other teams what seems to work in their sessions. However, JUNAEB does not systematically and frequently monitor each team's delivery.
Of the three teams we interviewed, only one had a workshop observed over the last two years.\footnote{SFL employees also do not have monetary or non-monetary incentives tied to the quality of their workshops.} Then, SFL's lack of effect may be due to the lack of a frequent and intensive third-party monitoring of the workshops, unlike what is happening in the studies reviewed by \cite{payton2008} and \cite{wigelsworth2016impact}. Without sufficient monitoring, teams may not implement the program with high-enough fidelity.
Unfortunately, beyond the striking difference between SFL and the reviewed interventions on that dimension, we cannot further support that conjecture by testing whether the treatment effect is larger for teams that are monitored more often or that implement SFL with higher fidelity. We do not observe the frequency at which each municipal team in our study is monitored by JUNAEB, and in any case our discussions with the teams and JUNAEB officials suggest monitoring is weak in every town. Similarly, we do not have a measure of implementers' fidelity and we do not observe implementers' number of years of experience into the program, which could be a proxy for fidelity.}

\section{Conclusion}
\label{sec:conclusion}

We explore the effects of ``Skills for life'' (SFL), a nationwide school-based SEL program for disruptive second graders in Chile. Eligibility to the program is based on first-grade teachers' ratings of students' disruptiveness, and SFL workshops consist in 10 two-hours sessions during which psychologists help students recognize and express their emotions, and teach them techniques to improve their behavior. We randomly assigned 172 classes to either receive SFL in the first or in the second semester of the 2015 school year, and we measured outcomes between the two semesters. Eligible students in treated classes see no improvement in their emotional stability, disruptiveness, and test scores. This is at odds with a large literature that has found large effects of SEL programs (see \citep{payton2008}, \citep{durlak2011impact}, \citep{dymnicki2012adolescent}, \citep{sklad2012effectiveness}, \citep{wigelsworth2016impact}, \citep{taylor2017promoting}, and \citep{corcoran2018effective} for recent meta-analyses).

\medskip
To understand this discrepancy, we investigate the differences between SFL and the programs studied in the literature. First, we find that SFL is not less intensive than those other programs. Second, its population may be harder to treat: all the programs studied in the literature take place in high-income countries, where the prevalence of ADHD, a disorder correlated with conduct problems, is much lower than in Chile. Accordingly, each SFL workshop is more likely to comprise one or two very disruptive students than an SEL workshop in a high-income country, and the presence of those hard-to-treat students may lower the workshop's effectiveness. We actually find evidence that SFL may increase students' disruptiveness in classes that have at least one very disruptive eligible student. The mechanism seems to be that SFL increases the friendships between very disruptive and other eligible students. Then, very disruptive students may have a negative influence on those other eligible students. To remediate this, SFL could exclude very disruptive students from its workshops, and offer them another type of treatment, for instance one-on-one sessions with a psychologist. Third, the literature has only considered small-scale programs mounted by researchers or NGOs, and either delivered by the researchers or NGO personnel, or by personnel closely monitored by them. On the other hand, SFL is a governmental program, and the government does not monitor the workshops' content and quality. Most of the interventions reviewed in the literature are implemented at a very small scale, in a handful of schools: the median number of treated students in the interventions reviewed by \cite{payton2008} is equal to 36. On the other hand, SFL treats around 8,500 students per year, in thousands of schools. Monitoring SFL as intensively as the small-scale interventions in \cite{payton2008} would probably be costly, but our results suggests this may be worth trying.



\newpage
\bibliography{biblio}
\clearpage

\newpage
\huge
\begin{center}
\textnormal{For Online Publication}
\end{center}
\normalsize

\begin{appendices}
\renewcommand{\thetable}{A\arabic{table}}
\setcounter{table}{0}

\section{Tables}

\begin{table}[htbp] \centering
  \begin{threeparttable}
\caption{Baseline - endline correlations in the control group}
\label{tab:corr_1}
{\normalsize
\begin{tabular}{lccc}
\toprule\toprule
{}&{Correlation}&{P-value}&{N} \tabularnewline
&{(1)}&{(2)}&{(3)} \tabularnewline
\midrule
\midrule\multicolumn{4}{c}{Panel A: student-level measures}\tabularnewline\midrule
School happiness score&0.221&0.000&1735 \tabularnewline
Self-control score&0.141&0.000&1816 \tabularnewline
Self-esteem score&0.134&0.000&1841 \tabularnewline
Disruptiveness, teacher&0.419&0.000&1782 \tabularnewline
Disruptiveness, enumerator&0.126&0.000&1871 \tabularnewline
\% school days missed&0.033&0.084&2751 \tabularnewline
Spanish test score&0.525&0.000&1897 \tabularnewline
Math test score&0.508&0.000&1897 \tabularnewline
\% class friends with student&0.324&0.000&2245 \tabularnewline
Friends' average ability&0.408&0.000&1644 \tabularnewline
Friends' average disruptiveness&0.349&0.000&1502 \tabularnewline
No friends in the class&0.099&0.000&2245 \tabularnewline
\midrule\multicolumn{4}{c}{Panel B: class-level measures}\tabularnewline\midrule
Disruptiveness, teacher&0.5&0.000&78 \tabularnewline
Bullying in class, teacher&0.392&0.000&76 \tabularnewline
Disruptiveness, enumerator&0.254&0.024&79 \tabularnewline
Average decibels during class&0.152&0.18&79 \tabularnewline
Delay in class's start (minutes)&0.031&0.788&79 \tabularnewline
\bottomrule

\end{tabular}
\begin{tablenotes}
\item
\footnotesize {\it Notes:} This table reports the correlation, in control classes, of several covariates between baseline and endline. Column (1) reports the baseline - endline correlation of the covariates. Column (2) reports the p-value of the significance of the correlation. Column (3) reports the number of observations used to compute the correlation.
\end{tablenotes}
}
  \end{threeparttable}
\end{table}

\begin{table}[htbp] \centering
  \begin{threeparttable}
\caption{Correlations between baseline disruptiveness measures}
\label{tab:corr_2}
{\normalsize
\begin{tabular}{lccc}
\toprule\toprule
{}&{Correlation}&{P-value}&{N} \tabularnewline
&{(1)}&{(2)}&{(3)} \tabularnewline
\midrule
\midrule\multicolumn{4}{c}{Panel A: student-level measures}\tabularnewline\midrule
Enumerator 1 - enumerator 2&0.504&0.000&4075 \tabularnewline
Teacher - enumerator&0.293&0.000&4035 \tabularnewline
Teacher dis. - avg. test score&-0.277&0.000&4139 \tabularnewline
Enumerator dis. - avg. test score&-0.17&0.000&4594 \tabularnewline
\midrule\multicolumn{4}{c}{Panel B: class-level measures}\tabularnewline\midrule
Enumerator 1 - Enumerator 2&0.618&0.000&157 \tabularnewline
Enumerator - Teacher&0.337&0.000&159 \tabularnewline
Enumerator - decibels&0.2&0.011&163 \tabularnewline
Teacher - decibels&-0.018&0.82&157 \tabularnewline
\bottomrule
\end{tabular}
\begin{tablenotes}
\item
\footnotesize {\it Notes:} This table reports the correlation, in control classes, between several baseline measures of disruption. Column (1) reports the correlation between the measures. Column (2) reports the p-value of the significance of the correlation. Column (3) reports the number of observations used to compute the correlation.
\end{tablenotes}
}
  \end{threeparttable}
\end{table}

\begin{table}[htbp] \centering
\begin{threeparttable}
\caption{Characteristics of takers and non-takers}
\label{tab:statdes_takersnontakers1}
{\normalsize
\begin{tabular}{lccccccc}
\toprule\toprule
{}&{Non-takers}&{Takers}&{P-value}&{N} \tabularnewline
&{(1)}&{(2)}&{(3)}&{(4)} \tabularnewline
\midrule\multicolumn{5}{c}{Panel A: demographic characteristics}\tabularnewline\midrule
Male&0.667&0.567&0.05&655 \tabularnewline
Teen mother&0.415&0.368&0.43&525 \tabularnewline
Student lives with father&0.515&0.551&0.577&478 \tabularnewline
$\leq$ p20 social security score&0.842&0.741&0.016&596 \tabularnewline
$\leq$ p5 social security score&0.463&0.441&0.693&596 \tabularnewline
Mother's education&8.448&8.327&0.798&576 \tabularnewline
Father's education&8.014&8.198&0.727&485 \tabularnewline
\midrule\multicolumn{5}{c}{Panel B: baseline measures}\tabularnewline\midrule
School happiness score&0.08&-0.034&0.41&477 \tabularnewline
Self-control score&-0.27&-0.172&0.493&511 \tabularnewline
Self-esteem score&-0.233&-0.176&0.708&513 \tabularnewline
Overall disruptiveness TOCA&1.128&0.81&0.011&645 \tabularnewline
Disruptiveness, enumerator&0.7&0.397&0.051&517 \tabularnewline
Spanish test score&-0.496&-0.326&0.22&548 \tabularnewline
Math test score&-0.489&-0.248&0.085&548 \tabularnewline
\% class friends with student&0.069&0.079&0.168&539 \tabularnewline
Friends' average disruptiveness&0.324&0.241&0.604&422 \tabularnewline
\bottomrule
 \end{tabular}

  \begin{tablenotes}
\item
\footnotesize {\it Notes:} This table reports descriptive statistics for eligible students, comparing those who attended and did not attend the workshops. Column (1) reports the mean of the outcome variable for eligible students who did not attend any session. Column (2) reports the mean of the variable for eligible students who attended at least one session. Column (3) reports the p-value of a test that the two means are equal. Column (4) reports the number of observations used in the comparison.
\end{tablenotes}
}
  \end{threeparttable}
  \end{table}

\begin{table}[htbp] \centering
\begin{threeparttable}
\caption{Characteristics of teachers}
\label{tab:teachstatdes}
{\normalsize
\begin{tabular}{lccccccc}

\toprule\toprule

{}&{Mean}&{N} \tabularnewline
&{(1)}&{(2)} \tabularnewline
\midrule
Female&0.963&160 \tabularnewline
Age&42.78&159 \tabularnewline
University degree&0.863&160 \tabularnewline
Years of experience&16.547&161 \tabularnewline
Years of experience, school&8.568&162 \tabularnewline
\bottomrule
\end{tabular}
  \begin{tablenotes}
\item
\footnotesize {\it Notes:} This table reports descriptive statistics for teachers in the sample. Column (1) reports the mean of the variables and Column (2) reports the number of observations used to compute that mean.
\end{tablenotes}
}
  \end{threeparttable}
\end{table}

\begin{table}[htbp] \centering
\begin{threeparttable}
\caption{Test of differential attrition for eligible students}
\label{tab:H1}
{\normalsize
\begin{tabular}{lccccccc}

\toprule\toprule
{Variables}&{Control}&{T-C}&{S.E.}&{Unadj. P}&{Adj. P}&{N} \tabularnewline
&{(1)}&{(2)}&{(3)}&{(4)}&{(5)}&{(6)} \tabularnewline\midrule
Eligible students per class at endline&6.651&0.473&0.386&0.22&0.55&169 \tabularnewline
Join class btw baseline and endline&0.023&0.004&0.008&0.649&0.649&1229 \tabularnewline
In class at baseline and endline&0.941&0.024&0.014&0.078&0.389&1178 \tabularnewline
With all enumerators' measures&0.748&-0.035&0.03&0.247&0.308&1238 \tabularnewline
With teacher's disruption measure&0.768&-0.084&0.071&0.235&0.392&1238 \tabularnewline

\bottomrule
\end{tabular}
   \begin{tablenotes}
\item
\footnotesize {\it Notes:} This table reports results from OLS regressions of several dependent variables on a treatment indicator. For student-level dependent variables, the regression includes lottery fixed effects. For class-level dependent variables, the regression is computed with propensity score weights. Column (1) reports the mean of the outcome variable for the control group. Column (2) reports the coefficient of the treatment indicator. Column (3) reports the standard error of this coefficient, clustered at the lottery level. Column (4) reports the unadjusted p-value of this coefficient, while Column (5) reports its p-value adjusted for multiple testing, following the method proposed in \cite{benjamini1995controlling}. Finally, Column (6) reports the number of observations used in the regression. All the dependent variables were collected by the authors at endline.
\end{tablenotes}
}
  \end{threeparttable}
  \end{table}

\begin{table}[htbp] \centering
\begin{threeparttable}
\caption{Test of differential attrition for ineligible students}
\label{tab:H2}
{\normalsize
\begin{tabular}{lccccccc}

\toprule\toprule
{Variables}&{Control}&{T-C}&{S.E.}&{Unadj. P}&{Adj. P}&{N} \tabularnewline
&{(1)}&{(2)}&{(3)}&{(4)}&{(5)}&{(6)} \tabularnewline\midrule
Ineligible students per class at endline&25.518&-1.009&0.853&0.237&0.592&169 \tabularnewline
Join class btw baseline and endline&0.045&-0.005&0.008&0.553&0.691&4433 \tabularnewline
In class at baseline and endline&0.962&-0.001&0.007&0.842&0.842&4159 \tabularnewline
With all enumerators' measures&0.783&-0.048&0.027&0.074&0.371&4466 \tabularnewline
With teacher's disruption measure&0.753&-0.059&0.067&0.383&0.638&4466 \tabularnewline

\bottomrule

\end{tabular}
   \begin{tablenotes}
\item
\footnotesize {\it Notes:} This table reports results from OLS regressions of several dependent variables on a treatment indicator. For student-level dependent variables, the regression includes lottery fixed effects. For class-level dependent variables, the regression is computed with propensity score weights. Column (1) reports the mean of the outcome variable for the control group. Column (2) reports the coefficient of the treatment indicator. Column (3) reports the standard error of this coefficient, clustered at the lottery level. Column (4) reports the unadjusted p-value of this coefficient, while Column (5) reports its p-value adjusted for multiple testing, following the method proposed in \cite{benjamini1995controlling}. Finally, Column (6) reports the number of observations used in the regression. All the dependent variables were collected by the authors at endline.
\end{tablenotes}
}
  \end{threeparttable}
  \end{table}

\begin{table}[htbp] \centering
\begin{threeparttable}

\caption{Balancing tests of eligible students' baseline characteristics}
\label{tab:H4}
{\normalsize
\begin{tabular}{lccccccc}

\toprule\toprule
{Variables}&{Control}&{T-C}&{S.E.}&{Unadj. P}&{Adj. P}&{N} \tabularnewline
&{(1)}&{(2)}&{(3)}&{(4)}&{(5)}&{(6)} \tabularnewline
\midrule\multicolumn{7}{c}{Panel A: demographic characteristics}\tabularnewline\midrule
Male&0.581&-0.004&0.046&0.937&0.97&1238 \tabularnewline
Teen mother&0.343&0.018&0.031&0.549&0.885&991 \tabularnewline
Student lives with father&0.563&-0.012&0.034&0.726&1&899 \tabularnewline
Social security score&5564.943&137.239&173.203&0.428&0.828&1124 \tabularnewline
Payment rate in health services&2.879&0.327&0.361&0.365&0.963&1122 \tabularnewline
Mother's education&8.813&-0.292&0.32&0.362&1&1080 \tabularnewline
Father's education&8.743&-0.565&0.38&0.137&0.995&913 \tabularnewline

\midrule\multicolumn{7}{c}{Panel B: TOCA and PSC scores}\tabularnewline\midrule
Authority Acceptance TOCA&1.027&-0.084&0.063&0.181&0.751&1223 \tabularnewline
Social Contact TOCA&0.842&-0.025&0.072&0.723&1&1223 \tabularnewline
Motiv. for Schooling TOCA&0.842&-0.036&0.06&0.543&0.985&1223 \tabularnewline
Emotional Maturity TOCA&0.563&-0.12&0.076&0.117&1&1223 \tabularnewline
Atention and Focus TOCA&0.834&-0.054&0.063&0.391&0.873&1223 \tabularnewline
Activity Level TOCA&0.831&-0.054&0.064&0.404&0.837&1223 \tabularnewline
Academic ability TOCA&0.667&-0.016&0.071&0.82&0.951&1222 \tabularnewline
Overall disruptiveness TOCA&0.891&-0.046&0.076&0.548&0.935&1220 \tabularnewline
PSC&0.477&-0.011&0.08&0.889&0.955&903 \tabularnewline

\midrule\multicolumn{7}{c}{Panel C: baseline measures}\tabularnewline\midrule
School happiness score&-0.107&0.082&0.083&0.323&1&929 \tabularnewline
Self-control score&-0.148&-0.057&0.063&0.371&0.897&986 \tabularnewline
Self-esteem score&-0.107&-0.105&0.076&0.168&0.811&991 \tabularnewline
Disruptiveness, teacher&0.396&0.087&0.276&0.753&0.993&253 \tabularnewline
Disruptiveness, enumerator&0.242&0.204&0.085&0.017&0.484&1007 \tabularnewline
Spanish test score&-0.321&-0.021&0.086&0.806&0.973&1036 \tabularnewline
Math test score&-0.301&0.021&0.099&0.829&0.924&1036 \tabularnewline
\% class friends with student&0.075&0.002&0.006&0.769&0.97&1030 \tabularnewline
Friends' average ability&-0.09&-0.002&0.114&0.988&0.988&863 \tabularnewline
Friends' average disruptiveness&0.122&0.099&0.103&0.333&1&822 \tabularnewline
No friends in the class&0.128&0.047&0.026&0.065&0.938&1030 \tabularnewline
Distance to teacher's desk&4.361&-0.079&0.18&0.66&1&863 \tabularnewline
\% school days missed, March&36.971&-4.809&3.421&0.16&0.927&1236 \tabularnewline

\bottomrule
\end{tabular}
  \begin{tablenotes}[para,flushleft]
\item
\footnotesize {\it Notes:} This table reports results from OLS regressions of several dependent variables on a treatment indicator and lottery fixed effects for eligible students. Column (1) reports the mean of the outcome variable for the control group. Column (2) reports the coefficient of the treatment indicator. Column (3) reports the standard error of this coefficient, clustered at the lottery level. Column (4) reports the unadjusted p-value of this coefficient, while Column (5) reports its p-value adjusted for multiple testing, following the method proposed in \cite{benjamini1995controlling}. Finally, Column (6) reports the number of observations used in the regression.
\end{tablenotes}
}
\end{threeparttable}
\end{table}

\begin{table}[htbp] \centering
\begin{threeparttable}
\caption{Balancing tests of eligible students' baseline characteristics, for those with all enumerators' endline measures.}
\label{tab:H5}
{\normalsize
\begin{tabular}{lccccccc}

\toprule\toprule
{Variables}&{Control}&{T-C}&{S.E.}&{Unadj. P}&{Adj. P}&{N} \tabularnewline
&{(1)}&{(2)}&{(3)}&{(4)}&{(5)}&{(6)} \tabularnewline
\midrule\multicolumn{7}{c}{Panel A: demographic characteristics}\tabularnewline\midrule
Male&0.56&0.016&0.054&0.767&1&906 \tabularnewline
Teen mother&0.324&0.081&0.04&0.044&0.632&731 \tabularnewline
Student lives with father&0.58&-0.051&0.038&0.183&0.883&665 \tabularnewline
Social security score&5640.612&-62.531&227.803&0.784&1&819 \tabularnewline
Payment rate in health services&3.005&0.122&0.472&0.795&1&824 \tabularnewline
Mother's education&8.836&-0.218&0.404&0.589&1&794 \tabularnewline
Father's education&8.768&-0.197&0.396&0.619&1&667 \tabularnewline
\midrule\multicolumn{7}{c}{Panel B: TOCA and PSC scores}\tabularnewline\midrule
Authority Acceptance TOCA&1&-0.038&0.063&0.548&1&894 \tabularnewline
Social Contact TOCA&0.785&0.008&0.077&0.919&0.987&894 \tabularnewline
Motiv. for Schooling TOCA&0.809&-0.009&0.065&0.893&1&894 \tabularnewline
Emotional Maturity TOCA&0.591&-0.128&0.083&0.123&0.895&894 \tabularnewline
Atention and Focus TOCA&0.798&0.013&0.064&0.845&1&894 \tabularnewline
Activity Level TOCA&0.821&-0.026&0.07&0.713&1&894 \tabularnewline
Academic ability TOCA&0.626&-0.014&0.079&0.859&1&894 \tabularnewline
Overall disruptiveness TOCA&0.801&0.034&0.092&0.712&1&893 \tabularnewline
PSC&0.441&-0.005&0.089&0.957&0.991&669 \tabularnewline

\midrule\multicolumn{7}{c}{Panel C: baseline measures}\tabularnewline\midrule
School happiness score&-0.077&0.069&0.091&0.445&1&700 \tabularnewline
Self-control score&-0.136&-0.018&0.079&0.824&1&745 \tabularnewline
Self-esteem score&-0.13&-0.043&0.093&0.643&1&744 \tabularnewline
Disruptiveness, teacher&0.341&0.061&0.215&0.776&1&192 \tabularnewline
Disruptiveness, enumerator&0.201&0.203&0.095&0.033&0.957&742 \tabularnewline
Spanish test score&-0.264&-0.01&0.084&0.908&1&769 \tabularnewline
Math test score&-0.22&0.037&0.11&0.736&1&769 \tabularnewline
\% class friends with student&0.077&0.006&0.006&0.353&1&765 \tabularnewline
Friends' average ability&-0.071&0.000&0.126&0.997&0.997&656 \tabularnewline
Friends' average disruptiveness&0.094&0.162&0.118&0.17&0.987&623 \tabularnewline
No friends in the class&0.111&0.048&0.026&0.068&0.657&765 \tabularnewline
Distance to teacher's desk&4.377&-0.203&0.225&0.366&1&630 \tabularnewline
\% school days missed, March&37.887&-4.312&3.658&0.238&0.988&904 \tabularnewline

\bottomrule

\end{tabular}
  \begin{tablenotes}[para,flushleft]
\item
\footnotesize {\it Notes:} This table reports results from OLS regressions of several dependent variables on a treatment indicator and lottery fixed effects for eligible students with all enumerators' endline measures. Column (1) reports the mean of the outcome variable for the control group. Column (2) reports the coefficient of the treatment indicator. Column (3) reports the standard error of this coefficient, clustered at the lottery level. Column (4) reports the unadjusted p-value of this coefficient, while Column (5) reports its p-value adjusted for multiple testing, following the method proposed in \cite{benjamini1995controlling}. Finally, Column (6) reports the number of observations used in the regression.
\end{tablenotes}
}
\end{threeparttable}
\end{table}

\begin{table}[htbp] \centering
\begin{threeparttable}
\caption{Balancing tests of eligible students' baseline characteristics, for those with teacher's endline disruptiveness measure.}
\label{tab:H6}
{\normalsize
\begin{tabular}{lccccccc}

\toprule\toprule
{Variables}&{Control}&{T-C}&{S.E.}&{Unadj. P}&{Adj. P}&{N} \tabularnewline
&{(1)}&{(2)}&{(3)}&{(4)}&{(5)}&{(6)} \tabularnewline
\midrule\multicolumn{7}{c}{Panel A: demographic characteristics}\tabularnewline\midrule
Male&0.574&-0.01&0.053&0.848&0.984&901 \tabularnewline
Teen mother&0.337&0.033&0.038&0.394&0.952&724 \tabularnewline
Student lives with father&0.564&-0.006&0.045&0.89&0.922&659 \tabularnewline
Social security score&5533.873&205.674&236.641&0.385&1&814 \tabularnewline
Payment rate in health services&3.144&-0.045&0.506&0.929&0.929&816 \tabularnewline
Mother's education&8.897&-0.594&0.415&0.152&0.883&798 \tabularnewline
Father's education&8.771&-0.483&0.511&0.345&1&673 \tabularnewline

\midrule\multicolumn{7}{c}{Panel B: TOCA and PSC scores}\tabularnewline\midrule
Authority Acceptance TOCA&0.983&-0.12&0.08&0.136&0.983&889 \tabularnewline
Social Contact TOCA&0.829&0.041&0.096&0.666&1&889 \tabularnewline
Motiv. for Schooling TOCA&0.852&-0.018&0.081&0.821&0.992&889 \tabularnewline
Emotional Maturity TOCA&0.597&-0.123&0.1&0.219&1&889 \tabularnewline
Atention and Focus TOCA&0.842&-0.046&0.082&0.572&0.922&889 \tabularnewline
Activity Level TOCA&0.821&-0.124&0.081&0.124&1&889 \tabularnewline
Academic ability TOCA&0.676&-0.052&0.091&0.563&0.961&888 \tabularnewline
Overall disruptiveness TOCA&0.877&-0.069&0.099&0.482&0.999&887 \tabularnewline
PSC&0.434&-0.017&0.103&0.869&0.933&662 \tabularnewline

\midrule\multicolumn{7}{c}{Panel C: baseline measures}\tabularnewline\midrule
School happiness score&-0.064&-0.064&0.096&0.503&0.912&680 \tabularnewline
Self-control score&-0.128&-0.165&0.085&0.053&0.762&718 \tabularnewline
Self-esteem score&-0.078&-0.106&0.088&0.23&0.952&720 \tabularnewline
Disruptiveness, teacher&0.275&0.057&0.245&0.815&1&190 \tabularnewline
Disruptiveness, enumerator&0.193&0.107&0.105&0.31&1&743 \tabularnewline
Spanish test score&-0.34&0.03&0.088&0.736&1&758 \tabularnewline
Math test score&-0.28&0.036&0.133&0.786&1&758 \tabularnewline
\% class friends with student&0.075&0.006&0.008&0.451&1&751 \tabularnewline
Friends' average ability&-0.138&0.129&0.143&0.367&1&635 \tabularnewline
Friends' average disruptiveness&0.129&0.026&0.14&0.853&0.951&611 \tabularnewline
No friends in the class&0.102&0.088&0.035&0.011&0.31&751 \tabularnewline
Distance to teacher's desk&4.441&0.061&0.178&0.732&1&643 \tabularnewline
\% school days missed, March&37.204&-2.795&4.143&0.5&0.966&899 \tabularnewline

\bottomrule

\end{tabular}
  \begin{tablenotes}[para,flushleft]
\item
\footnotesize {\it Notes:} This table reports results from OLS regressions of several dependent variables on a treatment indicator and lottery fixed effects for eligible students with teacher's endline disruptiveness measure. Column (1) reports the mean of the outcome variable for the control group. Column (2) reports the coefficient of the treatment indicator. Column (3) reports the standard error of this coefficient, clustered at the lottery level. Column (4) reports the unadjusted p-value of this coefficient, while Column (5) reports its p-value adjusted for multiple testing, following the method proposed in \cite{benjamini1995controlling}. Finally, Column (6) reports the number of observations used in the regression.
\end{tablenotes}
}
\end{threeparttable}
\end{table}

\begin{table}[htbp] \centering
\begin{threeparttable}
\caption{Balancing tests of ineligible students' baseline characteristics.}
\label{tab:H7}
{\normalsize
\begin{tabular}{lccccccc}

\toprule\toprule
{Variables}&{Control}&{T-C}&{S.E.}&{Unadj. P}&{Adj. P}&{N} \tabularnewline
&{(1)}&{(2)}&{(3)}&{(4)}&{(5)}&{(6)} \tabularnewline
\midrule\multicolumn{7}{c}{Panel A: demographic characteristics}\tabularnewline\midrule
Male&0.486&0.026&0.027&0.327&0.678&4466 \tabularnewline
Teen mother&0.328&0.016&0.02&0.434&0.662&3449 \tabularnewline
Student lives with father&0.639&-0.012&0.017&0.501&0.727&2866 \tabularnewline
Social security score&5965.036&-108.938&107.006&0.309&0.746&3944 \tabularnewline
Payment rate in health services&4.132&-0.019&0.313&0.951&0.951&3927 \tabularnewline
Mother's education&9.239&-0.19&0.2&0.341&0.582&3647 \tabularnewline
Father's education&9.181&-0.017&0.177&0.925&0.958&3204 \tabularnewline

\midrule\multicolumn{7}{c}{Panel B: TOCA and PSC scores}\tabularnewline\midrule
Authority Acceptance TOCA&-0.356&0.059&0.054&0.278&1&3654 \tabularnewline
Social Contact TOCA&-0.346&0.14&0.055&0.01&0.297&3654 \tabularnewline
Motiv. for Schooling TOCA&-0.312&0.071&0.047&0.132&0.765&3654 \tabularnewline
Emotional Maturity TOCA&-0.171&0.024&0.092&0.795&0.922&3654 \tabularnewline
Atention and Focus TOCA&-0.32&0.092&0.053&0.086&0.624&3654 \tabularnewline
Activity Level TOCA&-0.33&0.124&0.066&0.059&0.86&3645 \tabularnewline
Academic ability TOCA&-0.244&0.043&0.041&0.292&0.847&3633 \tabularnewline
Overall disruptiveness TOCA&-0.335&0.075&0.041&0.068&0.66&3630 \tabularnewline
PSC&-0.171&0.043&0.044&0.333&0.603&2882 \tabularnewline

\midrule\multicolumn{7}{c}{Panel C: baseline measures}\tabularnewline\midrule
School happiness score&0.039&-0.015&0.039&0.697&0.879&3502 \tabularnewline
Self-control score&0.05&-0.005&0.045&0.917&0.985&3608 \tabularnewline
Self-esteem score&0.066&-0.051&0.043&0.234&0.971&3619 \tabularnewline
Disruptiveness, teacher&-0.132&0.052&0.181&0.772&0.933&804 \tabularnewline
Disruptiveness, enumerator&-0.067&0.078&0.06&0.193&0.933&3638 \tabularnewline
Spanish test score&0.139&-0.065&0.076&0.393&0.632&3722 \tabularnewline
Math test score&0.083&0.033&0.079&0.676&0.891&3722 \tabularnewline
\% class friends with student&0.09&-0.003&0.005&0.523&0.722&3691 \tabularnewline
Friends' average ability&0.055&0.017&0.099&0.86&0.959&3260 \tabularnewline
Friends' average disruptiveness&-0.094&0.075&0.073&0.305&0.804&3109 \tabularnewline
No friends in the class&0.097&0.02&0.02&0.328&0.635&3691 \tabularnewline
Distance to teacher's desk&4.519&0.168&0.158&0.286&0.923&3129 \tabularnewline
\% school days missed, March&38.922&-2.992&2.969&0.314&0.699&4427 \tabularnewline
\bottomrule

\end{tabular}
  \begin{tablenotes}[para,flushleft]
  \item
\footnotesize {\it Notes:} This table reports results from OLS regressions of several dependent variables on a treatment indicator and lottery fixed effects for ineligible students. Column (1) reports the mean of the outcome variable for the control group. Column (2) reports the coefficient of the treatment indicator. Column (3) reports the standard error of this coefficient, clustered at the lottery level. Column (4) reports the unadjusted p-value of this coefficient, while Column (5) reports its p-value adjusted for multiple testing, following the method proposed in \cite{benjamini1995controlling}. Finally, Column (6) reports the number of observations used in the regression.
\end{tablenotes}
}
\end{threeparttable}
\end{table}

\begin{table}[htbp] \centering
\begin{threeparttable}

\caption{Balancing tests of ineligible students' baseline characteristics, for those with all enumerators' endline measures.}
\label{tab:H8}
{\normalsize
\begin{tabular}{lccccccc}

\toprule\toprule
{Variables}&{Control}&{T-C}&{S.E.}&{Unadj. P}&{Adj. P}&{N} \tabularnewline
&{(1)}&{(2)}&{(3)}&{(4)}&{(5)}&{(6)} \tabularnewline
\midrule\multicolumn{7}{c}{Panel A: demographic characteristics}\tabularnewline\midrule
Male&0.473&0.038&0.027&0.154&0.64&3376 \tabularnewline
Teen mother&0.322&0.015&0.021&0.481&0.734&2646 \tabularnewline
Student lives with father&0.647&-0.008&0.021&0.702&0.783&2203 \tabularnewline
Social security score&5982.408&-99.568&119.473&0.405&0.903&2989 \tabularnewline
Payment rate in health services&4.305&-0.181&0.376&0.63&0.795&2974 \tabularnewline
Mother's education&9.239&-0.184&0.223&0.409&0.847&2788 \tabularnewline
Father's education&9.189&0.022&0.19&0.908&0.941&2454 \tabularnewline

\midrule\multicolumn{7}{c}{Panel B: TOCA and PSC scores}\tabularnewline\midrule
Authority Acceptance TOCA&-0.365&0.054&0.05&0.282&0.745&2768 \tabularnewline
Social Contact TOCA&-0.39&0.173&0.061&0.005&0.138&2768 \tabularnewline
Motiv. for Schooling TOCA&-0.351&0.074&0.05&0.137&0.661&2768 \tabularnewline
Emotional Maturity TOCA&-0.182&0.079&0.103&0.44&0.751&2768 \tabularnewline
Atention and Focus TOCA&-0.346&0.095&0.052&0.069&0.501&2768 \tabularnewline
Activity Level TOCA&-0.331&0.164&0.061&0.007&0.108&2762 \tabularnewline
Academic ability TOCA&-0.28&0.05&0.045&0.264&0.766&2759 \tabularnewline
Overall disruptiveness TOCA&-0.363&0.075&0.038&0.045&0.436&2756 \tabularnewline
PSC&-0.195&0.047&0.058&0.417&0.807&2210 \tabularnewline

\midrule\multicolumn{7}{c}{Panel C: baseline measures}\tabularnewline\midrule
School happiness score&0.045&-0.018&0.045&0.688&0.798&2715 \tabularnewline
Self-control score&0.07&-0.021&0.05&0.673&0.813&2789 \tabularnewline
Self-esteem score&0.102&-0.081&0.051&0.112&0.651&2797 \tabularnewline
Disruptiveness, teacher&-0.208&0.101&0.167&0.545&0.752&641 \tabularnewline
Disruptiveness, enumerator&-0.06&0.048&0.061&0.434&0.787&2805 \tabularnewline
Spanish test score&0.171&-0.067&0.071&0.347&0.838&2870 \tabularnewline
Math test score&0.106&0.042&0.08&0.598&0.789&2870 \tabularnewline
\% class friends with student&0.09&0.000&0.006&0.95&0.95&2852 \tabularnewline
Friends' average ability&0.073&0.012&0.099&0.904&0.971&2524 \tabularnewline
Friends' average disruptiveness&-0.095&0.101&0.075&0.176&0.636&2402 \tabularnewline
No friends in the class&0.098&0.014&0.022&0.515&0.746&2852 \tabularnewline
Distance to teacher's desk&4.522&0.124&0.163&0.446&0.718&2416 \tabularnewline
\% school days missed, March&38.416&-3.897&3.252&0.231&0.744&3353 \tabularnewline
\bottomrule

\end{tabular}
  \begin{tablenotes}[para,flushleft]
\item
\footnotesize {\it Notes:} This table reports results from OLS regressions of several dependent variables on a treatment indicator and lottery fixed effects for ineligible students with all enumerators' endline measures. Column (1) reports the mean of the outcome variable for the control group. Column (2) reports the coefficient of the treatment indicator. Column (3) reports the standard error of this coefficient, clustered at the lottery level. Column (4) reports the unadjusted p-value of this coefficient, while Column (5) reports its p-value adjusted for multiple testing, following the method proposed in \cite{benjamini1995controlling}. Finally, Column (6) reports the number of observations used in the regression.
\end{tablenotes}
}
\end{threeparttable}
\end{table}

\begin{table}[htbp] \centering
\begin{threeparttable}

\caption{Balancing tests of ineligible students' baseline characteristics, for those with teacher's endline disruptiveness measure.}
\label{tab:H9}
{\normalsize
\begin{tabular}{lccccccc}

\toprule\toprule
{Variables}&{Control}&{T-C}&{S.E.}&{Unadj. P}&{Adj. P}&{N} \tabularnewline
&{(1)}&{(2)}&{(3)}&{(4)}&{(5)}&{(6)} \tabularnewline
\midrule\multicolumn{7}{c}{Panel A: demographic characteristics}\tabularnewline\midrule
Male&0.486&0.061&0.03&0.043&0.248&3202 \tabularnewline
Teen mother&0.319&0.04&0.025&0.118&0.381&2490 \tabularnewline
Student lives with father&0.641&0.012&0.023&0.61&0.804&2071 \tabularnewline
Social security score&5966.787&18.269&149.837&0.903&1&2838 \tabularnewline
Payment rate in health services&4.271&-0.156&0.42&0.71&0.823&2826 \tabularnewline
Mother's education&9.281&-0.293&0.281&0.296&0.506&2637 \tabularnewline
Father's education&9.276&-0.151&0.272&0.579&0.8&2310 \tabularnewline

\midrule\multicolumn{7}{c}{Panel B: TOCA and PSC scores}\tabularnewline\midrule
Authority Acceptance TOCA&-0.347&0.056&0.067&0.405&0.652&2645 \tabularnewline
Social Contact TOCA&-0.378&0.24&0.075&0.001&0.041&2645 \tabularnewline
Motiv. for Schooling TOCA&-0.323&0.122&0.055&0.028&0.267&2645 \tabularnewline
Emotional Maturity TOCA&-0.136&0.012&0.116&0.915&0.948&2645 \tabularnewline
Atention and Focus TOCA&-0.329&0.121&0.055&0.027&0.393&2645 \tabularnewline
Activity Level TOCA&-0.308&0.082&0.074&0.27&0.56&2637 \tabularnewline
Academic ability TOCA&-0.245&0.06&0.049&0.222&0.536&2632 \tabularnewline
Overall disruptiveness TOCA&-0.328&0.104&0.048&0.032&0.229&2630 \tabularnewline
PSC&-0.172&0.075&0.06&0.212&0.558&2084 \tabularnewline

\midrule\multicolumn{7}{c}{Panel C: baseline measures}\tabularnewline\midrule
School happiness score&0.047&-0.061&0.05&0.227&0.506&2531 \tabularnewline
Self-control score&0.106&-0.117&0.058&0.046&0.22&2592 \tabularnewline
Self-esteem score&0.09&-0.107&0.063&0.089&0.367&2604 \tabularnewline
Disruptiveness, teacher&-0.268&0.285&0.172&0.097&0.353&634 \tabularnewline
Disruptiveness, enumerator&-0.095&0.083&0.065&0.201&0.582&2638 \tabularnewline
Spanish test score&0.118&0.009&0.078&0.906&0.973&2689 \tabularnewline
Math test score&0.094&0.059&0.101&0.56&0.813&2689 \tabularnewline
\% class friends with student&0.091&0.000&0.005&0.937&0.937&2659 \tabularnewline
Friends' average ability&0.045&0.058&0.123&0.635&0.767&2366 \tabularnewline
Friends' average disruptiveness&-0.073&0.096&0.091&0.289&0.523&2259 \tabularnewline
No friends in the class&0.088&0.023&0.021&0.277&0.536&2659 \tabularnewline
Distance to teacher's desk&4.565&0.158&0.226&0.483&0.738&2355 \tabularnewline
\% school days missed, March&39.314&-1.844&3.663&0.615&0.775&3178 \tabularnewline
\bottomrule

\end{tabular}

  \begin{tablenotes}[para,flushleft]
\item
\footnotesize {\it Notes:} This table reports results from OLS regressions of several dependent variables on a treatment indicator and lottery fixed effects for ineligible students with teacher's endline disruptiveness measure. Column (1) reports the mean of the outcome variable for the control group. Column (2) reports the coefficient of the treatment indicator. Column (3) reports the standard error of this coefficient, clustered at the lottery level. Column (4) reports the unadjusted p-value of this coefficient, while Column (5) reports its p-value adjusted for multiple testing, following the method proposed in \cite{benjamini1995controlling}. Finally, Column (6) reports the number of observations used in the regression.
\end{tablenotes}
}
\end{threeparttable}
\end{table}

\begin{table}[htbp] \centering
\begin{threeparttable}
\caption{Balancing tests of teachers' baseline characteristics}
\label{tab:H10}
{\normalsize
\begin{tabular}{lccccccc}

\toprule\toprule
{Variables}&{Control}&{T-C}&{S.E.}&{Unadj. P}&{Adj. P}&{N} \tabularnewline
&{(1)}&{(2)}&{(3)}&{(4)}&{(5)}&{(6)} \tabularnewline
\midrule\multicolumn{7}{c}{Panel A: demographic characteristics}\tabularnewline\midrule
Age&43.013&-0.256&1.763&0.885&0.965&159 \tabularnewline
University degree&0.872&-0.019&0.06&0.748&1&160 \tabularnewline
Years of experience&16.367&0.508&2.108&0.809&1&161 \tabularnewline
Years of experience in the school&8.139&0.729&1.331&0.584&1&162 \tabularnewline
Absenteeism&0.646&-0.101&0.547&0.853&1&162 \tabularnewline

\midrule\multicolumn{7}{c}{Panel B: motivation and taste for their job}\tabularnewline\midrule
Taste for her job&0.007&0.031&0.144&0.827&1&161 \tabularnewline
Confident to improve students' life&0.076&-0.146&0.172&0.395&1&161 \tabularnewline
Effort to prepare lectures&0.497&0.023&0.042&0.588&1&143 \tabularnewline
Diverse methods used in class&-0.005&0.016&0.161&0.919&0.919&161 \tabularnewline
\midrule\multicolumn{7}{c}{Panel C: mental health}\tabularnewline\midrule
Stress score&0.073&-0.138&0.156&0.377&1&160 \tabularnewline
Happiness score&0.148&-0.317&0.15&0.034&0.41&161 \tabularnewline
Control on life score&0.054&-0.115&0.151&0.447&1&158 \tabularnewline
\bottomrule

\end{tabular}
  \begin{tablenotes}[para,flushleft]
\item
\footnotesize {\it Notes:} This table reports results from OLS regressions of several dependent variables on a treatment indicator for teachers. The regression is estimated with propensity score weights. Column (1) reports the mean of the outcome variable for the control group. Column (2) reports the coefficient of the treatment indicator. Column (3) reports the standard error of this coefficient, clustered at the lottery level. Column (4) reports the unadjusted p-value of this coefficient, while Column (5) reports its p-value adjusted for multiple testing, following the method proposed in \cite{benjamini1995controlling}. Finally, Column (6) reports the number of observations used in the regression. All the dependent variables were collected by the authors at baseline.
\end{tablenotes}
}
\end{threeparttable}
\end{table}

\begin{table}[htbp] \centering
\begin{threeparttable}
\caption{Balancing tests of classes' baseline characteristics}
\label{tab:H12}
{\normalsize
\begin{tabular}{lccccccc}

\toprule\toprule
{Variables}&{Control}&{T-C}&{S.E.}&{Unadj. P}&{Adj. P}&{N} \tabularnewline
&{(1)}&{(2)}&{(3)}&{(4)}&{(5)}&{(6)} \tabularnewline\midrule
Academic level of the class, teacher&0.059&-0.086&0.14&0.538&0.538&162 \tabularnewline
Disruptiveness, teacher&-0.143&0.286&0.16&0.074&0.148&161 \tabularnewline
Bullying in class, teacher&0.033&-0.094&0.147&0.519&0.623&160 \tabularnewline
Disruptiveness, enumerator&-0.131&0.275&0.153&0.072&0.217&168 \tabularnewline
Delay in class's start (minutes)&8.802&1.122&1.253&0.37&0.555&166 \tabularnewline
Average decibels during class&0.053&1.796&0.745&0.016&0.095&165 \tabularnewline
\bottomrule

\end{tabular}
  \begin{tablenotes}[para,flushleft]
\item
\footnotesize {\it Notes:} This table reports results from OLS regressions of several dependent variables on a treatment indicator. The regression is estimated with propensity score weights. Column (1) reports the mean of the outcome variable for the control group. Column (2) reports the coefficient of the treatment indicator. Column (3) reports the standard error of this coefficient, clustered at the lottery level. Column (4) reports the unadjusted p-value of this coefficient, while Column (5) reports its p-value adjusted for multiple testing, following the method proposed in \cite{benjamini1995controlling}. Finally, Column (6) reports the number of observations used in the regression. All the dependent variables were collected by the authors at baseline.

\end{tablenotes}
}
\end{threeparttable}
\end{table}

\begin{table}[htbp] \centering
\begin{threeparttable}

\caption{Balancing tests of classes' baseline characteristics, for classes with all  teacher's or enumerators' endline measures.}
\label{tab:H12_03}
{\normalsize
\begin{tabular}{lccccccc}

\toprule\toprule

{Variables}&{Control}&{T-C}&{S.E.}&{Unadj. P}&{Adj. P}&{N} \tabularnewline
&{(1)}&{(2)}&{(3)}&{(4)}&{(5)}&{(6)} \tabularnewline
\midrule\multicolumn{7}{c}{Panel A: classes with all teacher's measures}\tabularnewline\midrule
Academic level of the class, teacher&0.052&-0.095&0.143&0.509&0.611&150 \tabularnewline
Disruptiveness, teacher&-0.145&0.326&0.17&0.055&0.332&149 \tabularnewline
Bullying in class, teacher&0.036&-0.099&0.158&0.532&0.532&148 \tabularnewline
\midrule\multicolumn{7}{c}{Panel B: classes with all enumerators' measures}\tabularnewline\midrule
Disruptiveness, enumerator&-0.136&0.277&0.152&0.068&0.205&155 \tabularnewline
Average decibels during class&-0.108&1.391&0.815&0.088&0.176&153 \tabularnewline
Delay in class's start (minutes)&8.885&1.424&1.412&0.313&0.469&153 \tabularnewline
\bottomrule

\end{tabular}
  \begin{tablenotes}
\item
\footnotesize
{\it Notes:} This table reports results from OLS regressions of several dependent variables on a treatment indicator for classes with all teacher's or enumerators' measures. The regression is estimated with propensity score weights. Column (1) reports the mean of the outcome variable for the control group. Column (2) reports the coefficient of the treatment indicator. Column (3) reports the standard error of this coefficient, clustered at the lottery level. Column (4) reports the unadjusted p-value of this coefficient, while Column (5) reports its p-value adjusted for multiple testing, following the method proposed in \cite{benjamini1995controlling}. Finally, Column (6) reports the number of observations used in the regression. All the dependent variables were collected by the authors at baseline.
\end{tablenotes}
}
  \end{threeparttable}
\end{table}

\clearpage
\renewcommand{\thetable}{B\arabic{table}}
\setcounter{table}{0}

\section{Results without controls}

\begin{table}[htbp] \centering
  \begin{threeparttable}

\caption{Treatment effect on eligible students}
\label{tab:H14X}
{\normalsize
\begin{tabular}{lccccccc}
\toprule\toprule
{Variables}&{Control}&{T-C}&{S.E.}&{Unadj. P}&{Adj. P}&{N} \tabularnewline
&{(1)}&{(2)}&{(3)}&{(4)}&{(5)}&{(6)} \tabularnewline
\midrule\multicolumn{7}{c}{Panel A: emotional stability}\tabularnewline\midrule
School happiness score&-0.107&0.136&0.082&0.097&0.292&876 \tabularnewline
Self-control score&-0.184&-0.04&0.09&0.654&0.654&880 \tabularnewline
Self-esteem score&-0.17&-0.107&0.081&0.183&0.275&903 \tabularnewline
Standardized Treatment Effect&0.015&-0.002&0.08&0.977&&915 \tabularnewline
\midrule\multicolumn{7}{c}{Panel B: disruptiveness}\tabularnewline\midrule
Disruptiveness, teacher&0.353&0.057&0.099&0.562&1&904 \tabularnewline
Disruptiveness, enumerator&0.157&0.017&0.083&0.842&0.842&948 \tabularnewline
Standardized Treatment Effect&-0.025&0.041&0.088&0.645&&1110 \tabularnewline
\midrule\multicolumn{7}{c}{Panel C: academic outcomes}\tabularnewline\midrule
\% school days missed&12.82&1.055&1.016&0.299&0.896&1236 \tabularnewline
Spanish test score&-0.308&-0.044&0.082&0.59&0.886&956 \tabularnewline
Math test score&-0.274&-0.006&0.081&0.946&0.946&956 \tabularnewline
Standardized Treatment Effect&0.011&-0.049&0.083&0.555&&1238 \tabularnewline

\midrule\multicolumn{7}{c}{Panel D: integration in the class network}\tabularnewline\midrule

\% class friends with student&0.07&0.008&0.005&0.118&0.472&1147 \tabularnewline
Friends' average ability&-0.061&-0.022&0.096&0.816&0.816&829 \tabularnewline
Friends' average disruptiveness&0.177&0.146&0.096&0.13&0.259&787 \tabularnewline
No friends in the class&0.27&-0.025&0.027&0.348&0.464&1147 \tabularnewline
Standardized Treatment Effect&-0.008&0.035&0.066&0.592&&1148 \tabularnewline
\bottomrule

\end{tabular}
 \begin{tablenotes}
\item
\footnotesize {\it Notes:} This table reports results from OLS regressions of several dependent variables on a treatment indicator and lottery fixed effects for eligible students. Column (1) reports the mean of the outcome variable for the control group. Column (2) reports the coefficient of the treatment indicator. Column (3) reports the standard error of this coefficient, clustered at the lottery level. Column (4) reports the unadjusted p-value of this coefficient, while Column (5) reports its p-value adjusted for multiple testing, following the method proposed in \cite{benjamini1995controlling}. Finally, Column (6) reports the number of observations used in the regression. All the dependent variables, except for {\it \% school days missed}, were collected by the authors at endline.
\end{tablenotes}
}
  \end{threeparttable}
\end{table}

\begin{table}[htbp] \centering
  \begin{threeparttable}

\caption{Treatment effect on ineligible students}
\label{tab:H21X}
{\normalsize
\begin{tabular}{lccccccc}
\toprule\toprule
{Variables}&{Control}&{T-C}&{S.E.}&{Unadj. P}&{Adj. P}&{N} \tabularnewline
&{(1)}&{(2)}&{(3)}&{(4)}&{(5)}&{(6)} \tabularnewline
\midrule\multicolumn{7}{c}{Panel A: emotional stability}\tabularnewline\midrule
School happiness score&0.026&-0.009&0.04&0.828&0.828&3360 \tabularnewline
Self-control score&0.097&-0.067&0.044&0.126&0.377&3404 \tabularnewline
Self-esteem score&0.084&-0.066&0.047&0.161&0.241&3446 \tabularnewline
Standardized Treatment Effect&0.027&-0.062&0.046&0.183&&3476 \tabularnewline

\midrule\multicolumn{7}{c}{Panel B: disruptiveness}\tabularnewline\midrule
Disruptiveness, teacher&-0.212&0.258&0.104&0.014&0.027&3203 \tabularnewline
Disruptiveness, enumerator&-0.046&0.02&0.042&0.637&0.637&3518 \tabularnewline
Standardized Treatment Effect&-0.051&0.107&0.069&0.122&&4033 \tabularnewline

\midrule\multicolumn{7}{c}{Panel C: academic outcomes}\tabularnewline\midrule
\% school days missed&13.089&0.331&0.742&0.656&0.656&4427 \tabularnewline
Spanish test score&0.128&-0.097&0.07&0.167&0.5&3517 \tabularnewline
Math test score&0.08&-0.035&0.065&0.589&0.884&3517 \tabularnewline
Standardized Treatment Effect&0.018&-0.038&0.058&0.515&&4452 \tabularnewline

\midrule\multicolumn{7}{c}{Panel D: integration in the class network}\tabularnewline\midrule
\% class friends with student&0.087&0.002&0.003&0.538&0.718&4168 \tabularnewline
Friends' average ability&0.027&-0.033&0.1&0.745&0.745&3342 \tabularnewline
Friends' average disruptiveness&-0.11&0.097&0.07&0.163&0.652&3176 \tabularnewline
No friends in the class&0.197&-0.018&0.013&0.175&0.349&4168 \tabularnewline
Standardized Treatment Effect&0.003&0.001&0.051&0.992&&4171 \tabularnewline
\bottomrule
\end{tabular}
\begin{tablenotes}
\item
\footnotesize {\it Notes:} This table reports results from OLS regressions of several dependent variables on a treatment indicator and lottery fixed effects for ineligible students. Column (1) reports the mean of the outcome variable for the control group. Column (2) reports the coefficient of the treatment indicator. Column (3) reports the standard error of this coefficient, clustered at the lottery level. Column (4) reports the unadjusted p-value of this coefficient, while Column (5) reports its p-value adjusted for multiple testing, following the method proposed in \cite{benjamini1995controlling}. Finally, Column (6) reports the number of observations used in the regression. All the dependent variables, except for {\it \% school days missed}, were collected by the authors at endline.
\end{tablenotes}
}
  \end{threeparttable}
\end{table}

\begin{table}[htbp] \centering
  \begin{threeparttable}
\caption{Treatment effect on classroom environment}
\label{tab:H25X}
{\normalsize
\begin{tabular}{lccccccc}
\toprule\toprule
{Variables}&{Control}&{T-C}&{S.E.}&{Unadj. P}&{Adj. P}&{N} \tabularnewline
&{(1)}&{(2)}&{(3)}&{(4)}&{(5)}&{(6)} \tabularnewline\midrule
Disruptiveness, teacher&-0.187&0.39&0.131&0.003&0.015&160 \tabularnewline
Bullying in class, teacher&-0.038&0.062&0.159&0.698&0.698&160 \tabularnewline
Disruptiveness, enumerator&-0.186&0.389&0.148&0.009&0.021&167 \tabularnewline
Delay in class's start (minutes)&9.938&1.204&1.046&0.25&0.312&160 \tabularnewline
Average decibels during class&0.022&0.681&0.487&0.162&0.27&169 \tabularnewline
Standardized Treatment Effect&-0.215&0.424&0.131&0.001&&169 \tabularnewline
\bottomrule
\end{tabular}
\begin{tablenotes}
\item
\footnotesize {\it Notes:} This table reports results from OLS regressions of several dependent variables on a treatment indicator. The regression is estimated with propensity score weights. Column (1) reports the mean of the outcome variable for the control group. Column (2) reports the coefficient of the treatment indicator. Column (3) reports the standard error of this coefficient, clustered at the lottery level. Column (4) reports the unadjusted p-value of this coefficient, while Column (5) reports its p-value adjusted for multiple testing, following the method proposed in \cite{benjamini1995controlling}. Finally, Column (6) reports the number of observations used in the regression. All the dependent variables were collected by the authors at endline.
\end{tablenotes}
}
  \end{threeparttable}
\end{table}

\clearpage
\renewcommand{\thetable}{C\arabic{table}}
\setcounter{table}{0}

\section{Robustness checks for classes with at least one very disruptive student}

\begin{table}[H] \centering
  \begin{threeparttable}
\caption{Treatment effect in classes with at least one very disruptive students, without controls}
\label{tab:verydis_notctrl}
{\normalsize
\begin{tabular}{lcccccc}
\toprule\toprule
{Variables}&{Control}&{T-C}&{S.E.}&{Unadj. P}&{Adj. P}&{N} \tabularnewline
&{(1)}&{(2)}&{(3)}&{(4)}&{(5)}&{(6)} \tabularnewline
\midrule\multicolumn{6}{c}{Panel A: Very disruptive eligible students}\tabularnewline\midrule
Disruptiveness, teacher	&	0.985	&	-0.325	&	0.355	&	0.359	&	0.611	&	86	\tabularnewline
Disruptiveness, enumerator	&	0.286	&	0.613	&	0.439	&	0.162	&	0.306	&	85	\tabularnewline
Spanish test score	&	-0.460	&	0.038	&	0.335	&	0.910	&	0.910	&	88	\tabularnewline
Math test score	&	-0.230	&	0.063	&	0.512	&	0.902	&	0.958	&	88	\tabularnewline
\% class friends with student	&	0.051	&	0.025	&	0.012	&	0.042	&	0.103	&	109	\tabularnewline
\midrule\multicolumn{6}{c}{Panel B: Not very disruptive eligible students}\tabularnewline\midrule
Disruptiveness, teacher	&	0.294	&	0.451	&	0.128	&	0.000	&	0.007	&	391	\tabularnewline
Disruptiveness, enumerator	&	0.162	&	0.103	&	0.127	&	0.417	&	0.644	&	393	\tabularnewline
Spanish test score	&	-0.349	&	-0.176	&	0.106	&	0.095	&	0.202	&	397	\tabularnewline
Math test score	&	-0.349	&	-0.092	&	0.167	&	0.581	&	0.823	&	397	\tabularnewline
Friends with $\geq1$ very dis.	&	0.065	&	0.075	&	0.030	&	0.012	&	0.049	&	397	\tabularnewline
\midrule\multicolumn{6}{c}{Panel C: Ineligible students}\tabularnewline\midrule
Disruptiveness, teacher	&	-0.205	&	0.509	&	0.185	&	0.006	&	0.034	&	1517	\tabularnewline
Disruptiveness, enumerator	&	-0.093	&	0.172	&	0.077	&	0.025	&	0.086	&	1576	\tabularnewline
Spanish test score	&	0.035	&	-0.053	&	0.141	&	0.707	&	0.858	&	1579	\tabularnewline
Math test score	&	0.115	&	-0.031	&	0.177	&	0.862	&	0.977	&	1579	\tabularnewline
Friends with $\geq1$ very dis.	&	0.067	&	0.015	&	0.028	&	0.584	&	0.764	&	1577	\tabularnewline
\midrule\multicolumn{6}{c}{Panel D: Class-level outcomes}\tabularnewline\midrule
Disruptiveness, teacher	&	-0.250	&	0.669	&	0.236	&	0.005	&	0.039	&	72	\tabularnewline
Disruptiveness, enumerator	&	-0.250	&	0.516	&	0.245	&	0.035	&	0.100	&	76	\tabularnewline
\bottomrule
\end{tabular}
  \begin{tablenotes}[para,flushleft]
\footnotesize {\it Notes:} This table reports results from OLS regressions of several dependent variables on a treatment indicator. To account for the fact the randomization is stratified, the regressions in Panels A, B, and C have lottery fixed effects, while in the regressions in Panel D we use propensity score reweighting. Column (1) reports the mean of the outcome variable for the control group. Column (2) reports the coefficient of the treatment indicator. Column (3) reports the standard error of this coefficient, clustered at the lottery level. Column (4) reports the unadjusted p-value of this coefficient. Finally, Column (5) reports the number of observations used in the regression. All the dependent variables were collected by the authors at endline.
\end{tablenotes}
}
  \end{threeparttable}
\end{table}

\begin{table}[H] \centering
  \begin{threeparttable}
\caption{Treatment effect in classes with at least one very disruptive students, with extra controls}
\label{tab:verydis_forcedctrl}
{\normalsize
\begin{tabular}{lcccccc}
\toprule\toprule
{Variables}&{Control}&{T-C}&{S.E.}&{Unadj. P}&{Adj. P}&{N} \tabularnewline
&{(1)}&{(2)}&{(3)}&{(4)}&{(5)}&{(6)} \tabularnewline
\midrule\multicolumn{6}{c}{Panel A: Very disruptive eligible students}\tabularnewline\midrule
Disruptiveness, teacher	&	0.985	&	-0.097	&	0.403	&	0.811	&	0.984	&	86	\tabularnewline
Disruptiveness, enumerator	&	0.286	&	0.226	&	0.500	&	0.652	&	1	&	85	\tabularnewline
Spanish test score	&	-0.460	&	0.109	&	0.354	&	0.759	&	0.992	&	88	\tabularnewline
Math test score	&	-0.230	&	-0.076	&	0.576	&	0.895	&	1	&	88	\tabularnewline
\% class friends with student	&	0.051	&	0.011	&	0.018	&	0.544	&	1	&	109	\tabularnewline
\midrule\multicolumn{6}{c}{Panel B: Not very disruptive eligible students}\tabularnewline\midrule
Disruptiveness, teacher	&	0.294	&	0.482	&	0.148	&	0.001	&	0.019	&	391	\tabularnewline
Disruptiveness, enumerator	&	0.162	&	0.074	&	0.130	&	0.567	&	0.963	&	393	\tabularnewline
Spanish test score	&	-0.349	&	-0.196	&	0.101	&	0.052	&	0.146	&	397	\tabularnewline
Math test score	&	-0.349	&	-0.002	&	0.176	&	0.991	&	0.991	&	397	\tabularnewline
Friends with $\geq1$ very dis.	&	0.065	&	0.075	&	0.032	&	0.019	&	0.108	&	397	\tabularnewline
\midrule\multicolumn{6}{c}{Panel C: Ineligible students}\tabularnewline\midrule
Disruptiveness, teacher	&	-0.205	&	0.476	&	0.183	&	0.009	&	0.078	&	1517	\tabularnewline
Disruptiveness, enumerator	&	-0.093	&	0.122	&	0.082	&	0.136	&	0.331	&	1576	\tabularnewline
Spanish test score	&	0.035	&	0.012	&	0.124	&	0.922	&	0.979	&	1579	\tabularnewline
Math test score	&	0.115	&	0.057	&	0.152	&	0.709	&	1	&	1579	\tabularnewline
Friends with $\geq1$ very dis.	&	0.067	&	0.017	&	0.023	&	0.448	&	0.952	&	1577	\tabularnewline
\midrule\multicolumn{6}{c}{Panel D: Class-level outcomes}\tabularnewline\midrule
Disruptiveness, teacher	&	-0.250	&	0.543	&	0.261	&	0.038	&	0.161	&	72	\tabularnewline
Disruptiveness, enumerator	&	-0.250	&	0.492	&	0.246	&	0.045	&	0.153	&	76	\tabularnewline

\bottomrule
\end{tabular}
  \begin{tablenotes}[para,flushleft]
\footnotesize {\it Notes:} This table reports results from OLS regressions of several dependent variables on a treatment indicator and control variables. The control variables include those selected by a Lasso regression of the dependent variable on potential controls, following \cite{belloni2014high}, the variables unabalanced at baseline in the relevant subsample, and the baseline value of the outcome variable. To account for the fact the randomization is stratified, the regressions in Panels A, B, and C have lottery fixed effects, while in the regressions in Panel D we use propensity score reweighting. Column (1) reports the mean of the outcome variable for the control group. Column (2) reports the coefficient of the treatment indicator. Column (3) reports the standard error of this coefficient, clustered at the lottery level. Column (4) reports the unadjusted p-value of this coefficient. Finally, Column (5) reports the number of observations used in the regression. All the dependent variables were collected by the authors at endline.
\end{tablenotes}
}
  \end{threeparttable}
\end{table}

\end{appendices}
\end{document}